\documentclass[aps,twocolumn,pra,superscriptaddress,amsmath,amssymb]{revtex4-1} 
\usepackage{graphicx,hyperref}
\usepackage{color}
\usepackage{dcolumn}
\usepackage{comment}
\usepackage{bm}
\usepackage{txfonts}
\usepackage[T1]{fontenc}

\begin{document}

\title{Energy flow during relaxation in an electron-phonon system with multiple modes: A nonequilibrium Green's function study  }

\date{\today}

\author{Ken Inayoshi}
\email{k-inayoshi@stat.phys.titech.ac.jp}
\affiliation{
Department of Physics, Tokyo Institute of Technology, Meguro, Tokyo 152-8551, Japan
}

\author{Akihisa Koga}
\affiliation{
Department of Physics, Tokyo Institute of Technology, Meguro, Tokyo 152-8551, Japan
}

\author{Yuta Murakami}
\affiliation{
Center for Emergent Matter Science, RIKEN, Wako, Saitama 351-0198, Japan
}

\date{\today}

\begin{abstract}
We investigate an energy flow in an extended Holstein model
describing electron systems coupled to hot-phonons and heat-bath phonons.
To analyze the relaxation process after the photo-excitation of electrons,
we employ the nonequilibrium dynamical mean-field theory (DMFT). 
We find the backward energy flow during the relaxation,
where the direction of energy transfer between electrons and hot-phonons
is reversed.
To clarify the microscopic mechanism of the backward energy flow,
we introduce the approximated energy flows, which are calculated with
the gradient and quasiparticle approximations and are related to the nonequilibrium distribution functions.
We compare these approximated energy flows with the full energy flows calculated from the nonequilibrium DMFT. 
We find that, in the weak electron-hot-phonon coupling regime,
the full and approximated energy flows are almost the same,
meaning that 
the relaxation dynamics can be correctly understood
in terms of the nonequilibrium distribution functions. 
As the strength of the electron-hot-phonon coupling increases,
the approximated energy flow fails to qualitatively reproduce the full energy flow. 
This indicates that the microscopic origin of the energy flow cannot be solely explained by the nonequilibrium distribution functions. 
By comparing the energy flows with different levels of approximation, we reveal the role of the gradient and quasiparticle approximations.
\end{abstract}

\maketitle

\section{Introduction}
Understanding the relaxation dynamics of correlated electronic systems is a fundamental and essential objective in nonequilibrium condensed matter physics~\cite{RevModPhys.86.779,Giannetti2016}.
In electron-phonon coupled systems, the excess energy supplied to electrons by light is transferred throughout the system by interactions between electrons and between electrons and phonons.
The two-temperature model (2TM)~\cite{PhysRevLett.59.1460} has been used as the simple and effective method to describe the relaxation dynamics.
In this model, electrons and phonons are assumed to follow thermal distributions governed by time-dependent effective temperatures (the quasiequilibrium approximation), and the energy flow is proportional to the difference in these effective temperatures at each moment. 
However, recent theoretical and experimental studies have highlighted the significance of considering systems where not all phonon modes are governed by a single effective temperature as in the 2TM, for the detailed understanding  of the relaxation dynamics and the role of couplings between different degrees of freedom~\cite{ShotaOno2019,doi:10.1080/23746149.2022.2095925,CAPPELLUTI2022100664}.
For example, in some semimetals and semiconductors, the electrons around the band valley region emit only the phonons with a specific momentum~\cite{PRICE1985255,KOCEVAR1985155,PhysRevB.36.5016,PhysRevB.39.1180,PhysRevB.43.4158,PhysRevB.54.14487,doi:10.1063/1.363496,Butscher2007,PhysRevB.77.121402,doi:10.1063/1.3117236,PhysRevB.80.121403,doi:10.1063/1.3291615,PhysRevLett.104.227401,PhysRevB.83.153410,HUANG20111657,Scheuch2011,Wu2012,CAPPELLUTI2022100664}.
In metals, when the strength of the electron-phonon coupling varies significantly for each phonon mode, certain phonon modes that strongly couple with electrons quickly receive energy from electrons~\cite{PhysRevX.6.021003,PhysRevB.96.174439,PhysRevLett.119.097002,PhysRevLett.124.077001,PhysRevB.101.100302,SciPostPhys.12.5.173,doi:10.1063/1.4982738,CAPPELLUTI2022100664}.
Namely, there are modes that rapidly heat up after the excitation of electrons ("hot-phonon") and those that do not ("cold-phonon"), which should be treated as distinct subsystems~\cite{CAPPELLUTI2022100664}.

The multi-temperature model (MTM)~\cite{PhysRevX.6.021003,PhysRevB.96.174439,PhysRevB.98.134309,PhysRevB.101.035128,PhysRevLett.124.077001,PhysRevB.101.100302,SciPostPhys.12.5.173} is a natural extension of the 2TM to describe the relaxation dynamics in the system composed of multiple subsystems, such as electrons, hot-phonons and cold-phonons.
To avoid the arbitrariness of the fixed number of effective temperatures and the quasiequilibrium approximation in the MTM, the Boltzmann equation (BE) has also been actively used in the analysis of relaxation dynamics~\cite{PhysRevB.97.054310,ShotaOno2019,PhysRevB.101.100302,PhysRevB.102.184307,PhysRevB.103.125412}. 
These previous studies using the MTM and the BE have revealed the relaxation processes that cannot be explained by the 2TM.
One interesting example is the backward energy flow during the relaxation process.
Namely, the energy is transferred from photo-excited electrons to phonons in the initial relaxation process, but as time evolves, the energy flows from hot-phonons back to electrons~\cite{PhysRevB.97.054310,ShotaOno2019,PhysRevB.101.100302,PhysRevB.103.125412}.
Within the MTM, the mechanism of this backward energy flow can be intuitively understood as a reversal in the relative magnitude of the effective temperatures of electrons and hot-phonons. 

While the BE and the MTM offer intuitive insight into relaxation dynamics, it remains unclear whether these methods can properly explain relaxation dynamics.
This is because, in these methods, many approximations, such as the gradient, quasiparticle, and quasiequilibrium approximations,
are applied to the more microscopic theory, i.e., the nonequilibrium Green's function method (NEGF)~\cite{KadanoffBaym,Mahan2000,10.1143/PTP.123.581,stefanucci_vanleeuwen_2013,kamenev_2023,stefanucci2023in}.
Therefore, it is important to reveal the role of the effects ignored in these approximations and to identify the necessary elements for describing the relaxation dynamics.
One approach to achieve this objective is to compare the full energy flow calculated using the NEGF with the approximated energy flow obtained by sequentially applying the aforementioned approximations. 
In recent years, the NEGF has been intensively applied to
the Holstein model where electrons are coupled to a single phonon mode to examine the detailed relaxation process
and to verify the validity of the MTM and the BE~\cite{PhysRevB.88.165108,PhysRevX.3.041033,PhysRevB.87.235139,PhysRevB.90.075126,PhysRevB.91.045128,PhysRevX.8.041009,PhysRevB.98.245110,FREERICKS2021147104,picano2022stochastic}. 
On the other hand, it is not so trivial how applicable the approximations used in the BE and the MTM are
in the system composed of multiple subsystems where complex relaxation processes such as the backward energy flow should occur.

In this paper, we investigate the energy flow during the relaxation dynamics in an extended Holstein model. 
This model describes electron system coupled to both hot-phonons and heat-bath phonons.
We then apply the NEGF implemented with the nonequilibrium dynamical mean-field theory (DMFT)~\cite{https://doi.org/10.48550/arxiv.cond-mat/0202046,PhysRevLett.97.266408,RevModPhys.86.779,PhysRevB.88.165115,Turkowski2021} to the system.
In order to reveal the origin of the energy flow between subsystems, we introduce approximated energy flow equations extending methods used in previous studies~\cite{e18050180,https://doi.org/10.1002/prop.201600042,PhysRevX.8.041009,FREERICKS2021147104} and  employing the approximations used in the BE and the MTM. 
We compare the approximated energy flow with the full energy flow derived from the NEGF.
In the weak electron-hot-phonon coupling regime, we find that the approximated energy flow ruled by the nonequilibrium distribution functions qualitatively reproduces the full energy flow.
In addition, we show that the microscopic origin of the backward energy flow is explained by the reversal in the magnitude relation between the nonequilibrium distribution function (or the effective temperature) of the Green's function and that of the self-energy.
On the other hand, increasing the strength of the electron-hot-phonon coupling, we find that the approximated energy flow does not reproduce the full energy flow, and the effects ignored in the approximated energy flow become crucial.

This paper is organized as follows. 
In Sec.~\ref{sec:modelmethod}, we introduce the extended Holstein model and explain the nonequilibrium DMFT formalism.
In particular, we derive the energy flow equations based on the Kadanoff-Baym equation~\cite{KadanoffBaym}.
We also introduce the approximated energy flow equations based on the nonequilibrium distribution functions of the Green's functions and the self-energies.
In Sec.~\ref{sec:results}, we present the results of numerical simulations.
We compare the approximated energy flow with the full energy flow derived from the NEGF, and discuss the microscopic origin of the energy flow.
Sec.~\ref{sec:summary} provides a summary and an outlook.

\section{Model and Method} \label{sec:modelmethod}
\subsection{Hamiltonian and nonequilibrium DMFT}
\begin{figure}
\centering
\includegraphics[scale=0.4]{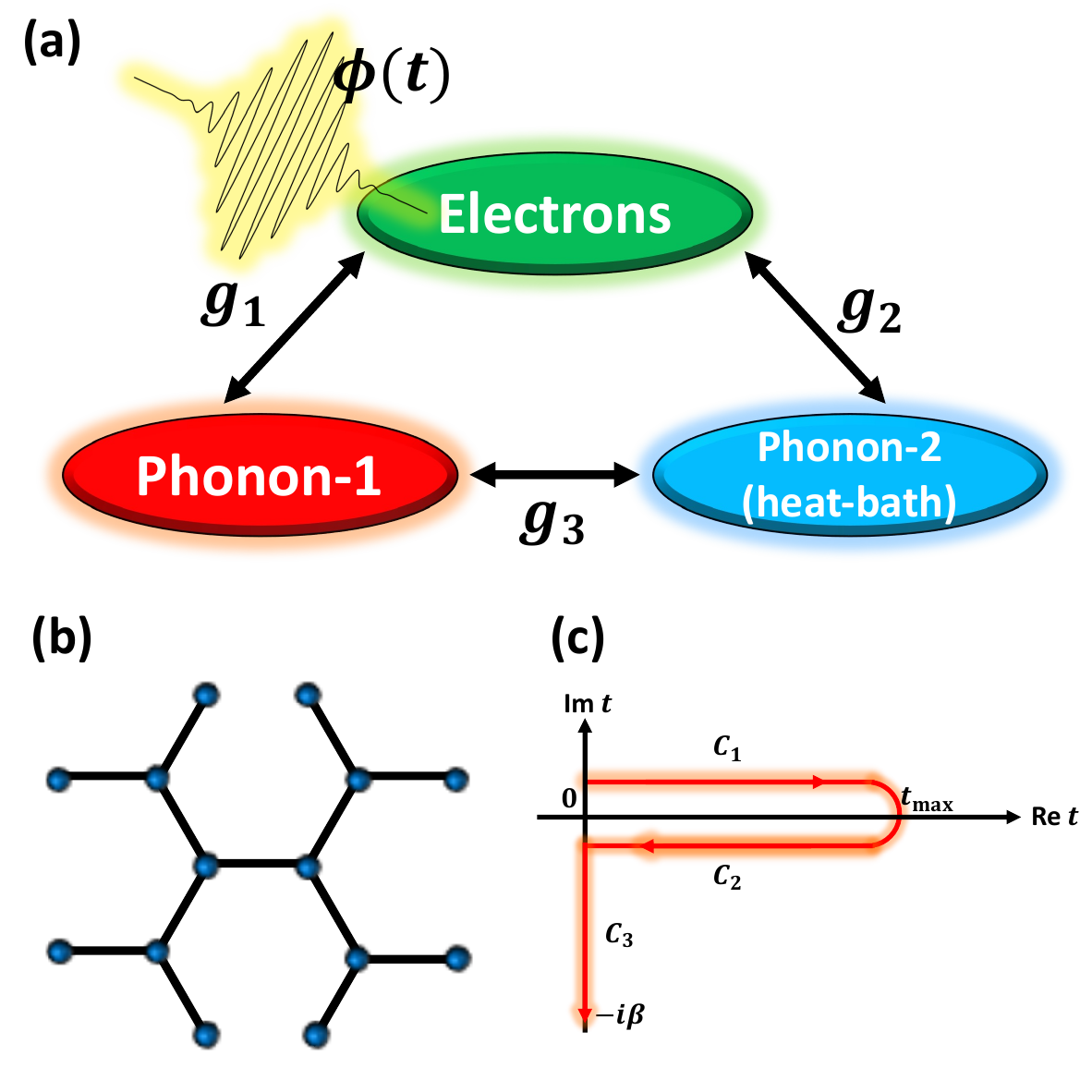}
\caption{(Color online) Schematic pictures of (a) the electro-phonon system in this paper, (b) the Bethe lattice (coordination number $z=3$), and (c) the Konstantinov-Perel's contour $\mathcal{C}$.}
\label{model}
\end{figure}

In this study, we consider an extended Holstein model where electrons couple to two types of phonon modes (Fig.~\ref{model}(a))~\cite{Covaci_2007}.
The Hamiltonian of the model is expressed as 
\begin{align}
\hat{H}(t)&=\sum_{\langle i,j\rangle\sigma}J_{ij}(t)\hat{c}_{i\sigma}^{\dagger}\hat{c}_{j\sigma} \nonumber \\
&+g_1\sum_i(\hat{n}_{i\uparrow}+\hat{n}_{i\downarrow}-1)(\hat{a}_i+\hat{a}_{i}^{\dagger})+\hbar\omega_1\sum_i\hat{a}_i^\dagger\hat{a}_i\nonumber \\
&+g_2\sum_{ip}(\hat{n}_{i\uparrow}+\hat{n}_{i\downarrow}-1)(\hat{b}_{ip}+\hat{b}_{ip}^{\dagger})+\sum_{ip}\hbar\omega_{2p}\hat{b}_{ip}^\dagger\hat{b}_{ip}\nonumber \\
&+g_3\sum_{ip}(\hat{a}_i+\hat{a}_{i}^{\dagger})(\hat{b}_{ip}+\hat{b}_{ip}^{\dagger}) \label{hamiltonian}.
\end{align}
Here, $\hat{c}_{i\sigma}\ (\hat{c}^{\dagger}_{i\sigma})$ is the annihilation (creation) operator of an electron with spin $\sigma$ on site $i$.
$\hat{a}\ (\hat{a}^{\dagger})$ is the annihilation (creation) operator of a phonon with frequency $\omega_1$ ("phonon-1").
$\hat{b}_p\ (\hat{b}^{\dagger}_p)$ is the annihilation (creation) operator of a phonon with frequency $\omega_{2p}$ and label $p$ ("phonon-2"). 
In this study, one particular phonon mode (phonon-1) is given the role as a hot-phonon, and the other phonons labeled with $p$ (phonon-2) are treated collectively as cold-phonons.
By considering a large number of electron-phonon couplings, the phonon-2 can be regarded as a heat-bath for the electronic subsystem.
$J_{ij}(t)$ is the time-dependent hopping parameter between nearest-neighbor sites,
$g_1$ ($g_2$) is the coupling between electrons and the phonon-1 (phonon-2), and 
$g_3$ is the Caldeira-Leggett coupling~\cite{PhysRevLett.46.211} between the phonon-1 and the phonon-2.
In this paper, for simplicity, we consider the model defined on the Bethe lattice with the infinite coordination number ($z\to \infty$) (Fig.~\ref{model}(b)),
where the DMFT is justified.
In the following, we focus on the half-filling condition of the electron number.

The different roles of the phonon-1 and the phonon-2 are implemented with the two types of the Migdal approximations~\cite{PhysRevB.91.045128,PhysRevB.98.245110}, as we discuss in detail in Sec.~\ref{subsec:impurity}.
In this work, we treat the phonon-1 as the hot-phonon, so the state of phonon-1 changes with time-evolution.
On the other hand, we treat the cold-phonon, i.e., phonon-2, as the heat-bath. 
Namely, the phonon-2 subsystem stays in equilibrium at the given temperature.
Strictly speaking, the state of phonon-2 can change with time-evolution by getting the energy from electrons.
Moreover, in realistic systems, the phonon energy dispersion depends on the momentum, and the energy transfer between phonons happens due to anharmonic phonon-phonon couplings in stead of the hybridization between phonons unlike in the above model~\cite{PhysRevB.96.174439,PhysRevB.97.054310,PhysRevB.102.184307,PhysRevB.103.125412}.  
However, the purpose of this study is to examine the general trend of the energy flow involving multiple electron-phonon couplings.
Therefore, we adopt the system containing minimal elements for this purpose.

In order to solve the problem, we use the nonequilibrium DMFT~\cite{https://doi.org/10.48550/arxiv.cond-mat/0202046,PhysRevLett.97.266408,RevModPhys.86.779,PhysRevB.88.165115,Turkowski2021} formulated on the L-shaped Konstantinov-Perel's contour $\mathcal{C}$~\cite{KonstantinovPerel1961,stefanucci_vanleeuwen_2013,Secchi_2019}. 
This contour consists of the forward real-time branch $\mathcal{C}_1$, the backward real-time branch $\mathcal{C}_2$, and the Matsubara imaginary-time branch $\mathcal{C}_3$ (see Fig.~\ref{model}(c)).
In the DMFT~\cite{RevModPhys.68.13,Turkowski2021}, the lattice self-energy is assumed to be local, and the local self-energy and the local lattice Green's function are identified with those of the effective impurity model, i.e., $\Sigma_{i,j}^{\rm latt}=\delta_{i,j}\Sigma^{\rm imp}$ and $G_{i,i}^{\rm latt}=G^{\rm imp}$ (this statement is exact in the infinite dimensional system).
We define the local Green's functions of electrons and phonon-1 on the contour $\mathcal{C}$ as
\begin{align}
G(t,t')&=-i\langle T_\mathcal{C}\hat{c}(t)\hat{c}^{\dagger}(t')\rangle, \\
D_1(t,t')&=-i\langle T_\mathcal{C}\hat{X}_{1}(t)\hat{X}_{1}(t')\rangle,
\end{align}
where $T_{\mathcal C}$ is the time ordering operator on the contour $\mathcal{C}$ and $\hat{X}_1=\hat{a}+\hat{a}^{\dagger}$ is the position operator for the phonon-1.
Assuming the homogeneous system, we omit the site index. 
The Green's functions $G$ and $D_1$ satisfy the Dyson equations of the impurity model,
\begin{align}
G(t,t')&=\mathcal{G}_0(t,t')+\left[ \mathcal{G}_0\ast \Sigma_{\rm el} \ast G\right](t,t'), \\
D_1(t,t')&=D_1^0(t,t')+\left[ D_1^0\ast \Pi_{\rm ph1} \ast D_1\right](t,t'),
\end{align}
where $\ast$ denotes the convolution on the contour $\mathcal{C}$.
$\Sigma_{\rm el}$  and $\Pi_{\rm ph1}$ represent the self-energy of the electron and that of the phonon-1 at the impurity site, respectively.
The Weiss Green's function $\mathcal{G}_0$ for the impurity problem satisfies  
\begin{align}
i\hbar\partial_t\mathcal{G}_0(t,t')-\left[\Delta \ast \mathcal{G}_0 \right](t,t')=\delta_\mathcal{C}(t,t'),
\end{align}
where $\delta_{\mathcal C}$ is the Dirac delta function on the contour $\mathcal{C}$ and $\Delta$ is the hybridization function.
The hybridization function is determined so that the impurity Green's function and self-energy match the lattice Green's function and self-energy (the DMFT self-consistency).
However, for the Bethe lattice, the expression of this self-consistency becomes simple as below.
The free phonon-1 Green's function $D_1^0$ is expressed as
\begin{align}
D_1^0(t,t')&=-i\left[ \theta_{\mathcal C}(t,t')+f_b(\omega_1)\right]e^{-i\omega_1(t-t')} \\ \nonumber
&-i\left[ \theta_{\mathcal C}(t',t)+f_b(\omega_1)\right]e^{i\omega_1(t-t')}.
\end{align}
Here, $\theta_{\mathcal C}$ is the Heaviside step function on the contour $\mathcal{C}$.
$f_b(\omega)=1/(\exp(\beta\omega)-1)$ is the Bose-Einstein distribution function.

We assume that the electron subsystem is directly photo-exited by the pump light.
The light-matter coupling is described with the Peierls substitution,
\begin{align}
J_{ij}(t)=
J_{ij}\exp\left\{ -\frac{ie}{\hbar}\int^{\bm{R}_i}_{\bm{R}_j}\bm{A}( t)\cdot \bm{dr} \right\},
\end{align}
where $-e$ ($e>0$) is the charge of an electron and $\bm{A}(t)$ is the vector potential. 
This vector potential is related to the electric field $\bm{E}(t)$ by $\bm{E}(t)=-\frac{\partial \bm{A}(t)}{\partial t}$.
Here, we assume the vector potential is uniform (the dipole approximation).
For the Bethe lattice with the infinite coordination number, the DMFT self-consistency including the effects of the Peierls substitution becomes as ~\cite{PhysRevB.95.195405}
\begin{align}
\Delta(t,t')&=J^{*}\cos\phi(t)G(t,t')J^{*}\cos\phi(t') \\ \nonumber
&+J^{*}\sin\phi(t)G(t,t')J^{*}\sin\phi(t').
\end{align}
Here, $J_{ij}=-J=-J^{*}/\sqrt{z}$ with renormalized hopping amplitude $J^{*}$ and $\phi(t)=eaA(t)$ with the lattice constant $a$.
$aA(t)$ corresponds to the projection of the vector potential $\bm{A}$ along the bond.
Previous numerical studies have confirmed that this setup on the Bethe lattice yields qualitatively the same results as the system on the hyper-cubic lattice with a field along the body-diagonal direction, see Ref.~\cite{Murakami2023arxiv} for example.

In the following, we set $\hbar=1$, $e=1$ and $a=1$.
We take $J^*$ and $1/J^*$ as the units of energy and time, respectively.
Thus, the electronic bandwidth for the noninteracting case becomes $W=4$.
Our time unit corresponds to the time scale of electron hopping, which is a few femtoseconds in typical materials.

\subsection{Impurity solvers}\label{subsec:impurity}
\begin{figure}
\centering
\includegraphics[scale=0.25]{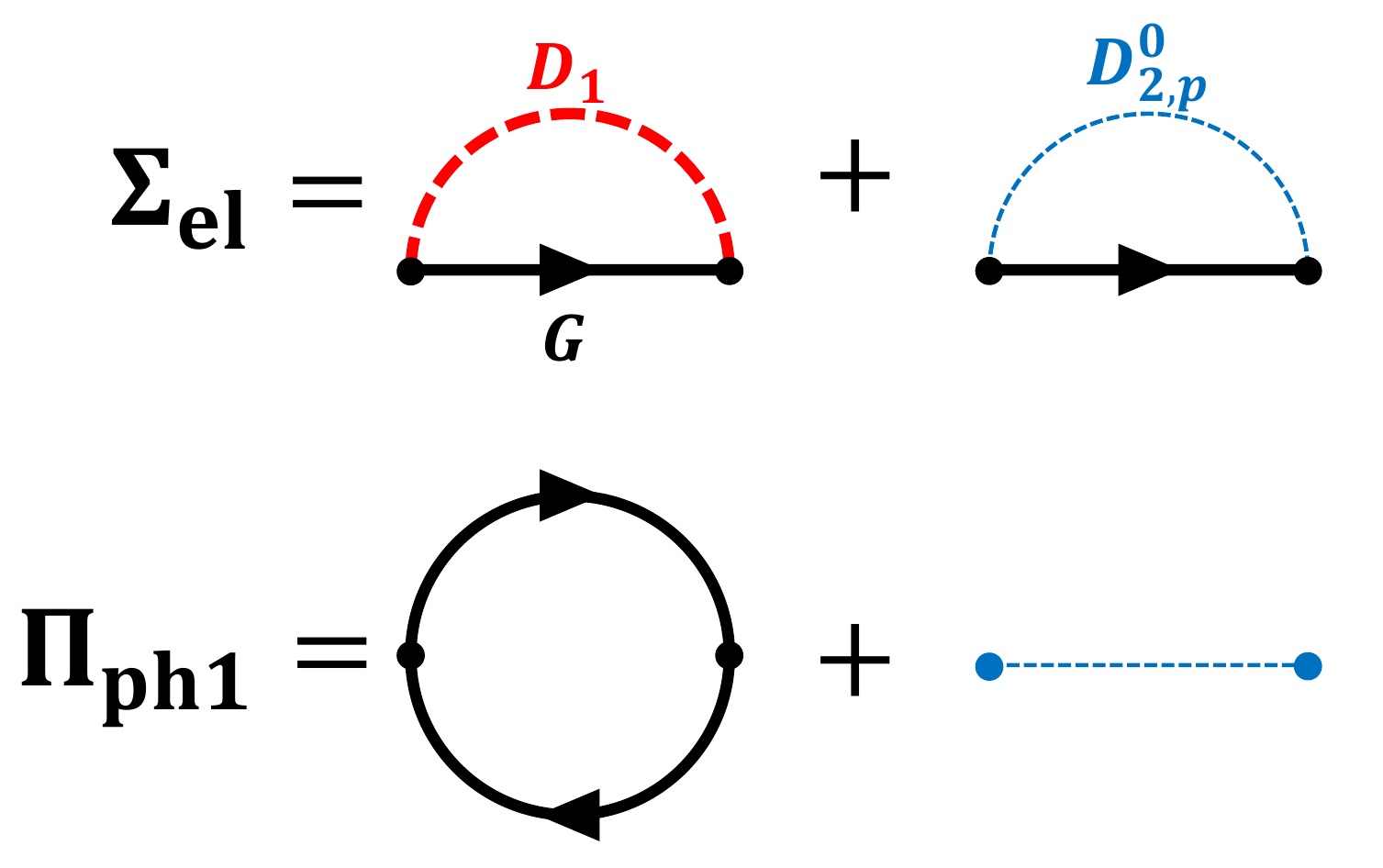}
\caption{(Color online) Self-energy diagrams of $\Sigma_{\rm el}$ and $\Pi_{\rm ph1}$.}
\label{FD}
\end{figure}

In this study, we focus on the weak electron-phonon coupling regime and use the Migdal approximation to solve the impurity model in the DMFT.
When the phonon energy is sufficiently small compared to the electron energy (energy band width), the Migdal approximation is justified because the vertex correlation of the self-energy can be neglected~\cite{Migdal1958,Eliashberg1960}. 
In the present case, since the roles of the phonon-1 and the phonon-2 are different, we apply the different Migdal approximations for them~\cite{PhysRevB.91.045128,PhysRevB.98.245110}.
We consider the situation where the phonon-1 and electrons dynamically affect each other. 
This situation can be captured by the self-consistent Migdal approximation, where we consider the self-energy of electrons and phonons in the self-consistent manner, as seen below. 
On the other hand, we regard the phonon-2 as the heat-bath. This situation can be captured by the unrenormalized Migdal approximation. 
The self-energies originating from electron-phonon couplings are expressed as
\begin{align}
\Sigma_{\rm el-ph1}(t,t')&=ig_1^2D_1(t,t')G(t,t'), \\
\Sigma_{\rm el-ph2}(t,t')&=i\left\{ g_2^2\sum_p D_{2p}^0(t,t')\right\}G(t,t').
\end{align}
The function $D_{2p}^0(t,t')$ is the free phonon-2 Green's function of the mode with label $p$~\cite{PhysRevB.96.045125}.
The total electron self-energy is
\begin{align}
\Sigma_{\rm el}(t,t')=\Sigma_{\rm el-ph1}(t,t')+\Sigma_{\rm el-ph2}(t,t').
\end{align}

In the self-consistent Migdal approximation, the phonon-1 self-energy originating from the electron-phonon-1 coupling is expressed as
\begin{align}
\Pi_{\rm el-ph1}(t,t')=-2ig_1^2G(t,t')G(t',t).
\end{align}
The self-energy originating from the coupling between the phonon-1 and the phonon-2 is expressed as
\begin{align}
\Pi_{\rm ph1-ph2}(t,t')=g_3^2\sum_p D_{2p}^0(t,t'). \label{eq:Pi_ph1ph2}
\end{align}
The total self-energy of phonon-1 is
\begin{align}
\Pi_{\rm ph1}(t,t')=\Pi_{\rm el-ph1}(t,t')+\Pi_{\rm ph1-ph2}(t,t').
\end{align}
The Feynman diagrams of above self-energies are shown in Fig.~\ref{FD}.
Since we use the unrenormalized Migdal approximation, we do not consider the self-energy of the phonon-2.

\subsection{Heat-bath phonon}
Now we explain the properties of the phonon-2 (heat-bath phonon).
In the Migdal approximations mentioned above, the effects of the phonon-2 show up through $\sum_p D_{2p}^0$, 
see the expressions of $\Sigma_{\rm el-ph2}$ and $\Pi_{\rm ph1-ph2}$.
Thus, only the form of  $\sum_p D_{2p}^0$ is relevant in our study.
This term can be expressed as 
\begin{align}
\sum_p D_{2p}^0(t,t') 
=-i\int^{\infty}_{-\infty}B(\omega)e^{-i\omega(t-t')}\left[ \theta_C(t,t')+f_b(\omega) \right] \ d\omega.
\end{align}
The bath spectrum $B(\omega)$ is defined as
\begin{align}
B(\omega)=\sum_p\left\{\delta(\omega-\omega_{2p})-\delta(\omega+\omega_{2p})\right\}.
\end{align}
In this study, we consider that this bath spectrum is expressed by the Cauchy-Lorentz distribution function with the location parameter $\omega_D$ and the half width $\gamma$~\cite{PhysRevB.96.045125},
\begin{align}
B(\omega)=\frac{1}{\pi}\left\{ \frac{\gamma}{(\omega-\omega_D)^2+\gamma^2}-\frac{\gamma}{(\omega+\omega_D)^2+\gamma^2} \right\}. \label{eq:lorentzDOS}
\end{align}

\subsection{Energy flow}
\subsubsection{Energies}
In this paper, we mainly analyze the time evolution of 
the electron kinetic energy $E_{\rm kin}$ and the phonon-1 energy $E_{\rm ph1}$ per site (the number of lattice sites is $N$).
They are expressed as ~\cite{PhysRevB.91.045128,PhysRevB.98.245110}
\begin{align}
E_{\rm kin}(t)&=\left\langle \frac{1}{N}\sum_{\langle i,j\rangle\sigma}J_{ij}(t)\hat{c}_{i\sigma}^{\dagger}(t)\hat{c}_{j\sigma}(t)\right\rangle \nonumber \\
&=-2i\left[ \Delta \ast G \right]^{<} (t,t), \label{eq:Ekin} \\ 
E_{\rm ph1}(t)&=\left\langle \frac{1}{N}\sum_i\omega_1\hat{a}_i^\dagger(t) \hat{a}_i(t) \right\rangle \nonumber \\
&=\frac{\omega_1}{4}\left[ iD_1^<(t,t)+\left. \frac{i\partial_t\partial_{t'}}{\omega_1^2}D_1^<(t,t')\right|_{t'=t}\right]-\frac{\omega_1}{2}. \label{eq:Eph1}
\end{align}
Note that besides above two energies, the electron-phonon-1 (electron-phonon-2) interaction energy $E_{\rm el-ph1}$ ($E_{\rm el-ph2}$) and phonon-phonon interaction energy $E_{\rm ph1-ph2}$ change with time in the NEGF (see the expressions of interaction energies in Appendix~\ref{app:interaction}).
On the other hand, in the BE and the MTM, the time-evolution of these interaction energies are not taken into account and the energy conservation holds only between the electron kinetic energy and the phonon energy without interaction energy (see Appendix~\ref{app:GFtoBE}).
In this paper, we compare the energy flow derived from the NEGF with that calculated using the approximations employed in the BE and the MTM. 
Therefore, we study the time-evolution of  the electron kinetic energy $E_{\rm kin}$ and the phonon-1 energy $E_{\rm ph1}$ which are also treated in the BE and the MTM.

\subsubsection{Energy flow equations}
In order to examine the energy transfer between three subsystems (electron, phonon-1, and phonon-2), we analyze the energy flow equations, i.e., the time derivative of $E_{\rm kin}$ and $E_{\rm ph1}$.
These equations can be derived from the Kadanoff-Baym equation~\cite{KadanoffBaym} and the Langreth rules~\cite{stefanucci_vanleeuwen_2013}.
They are expressed as 
\begin{align}
\frac{dE_{\rm kin}(t)}{dt}&=I_{\rm el}(t)+W_{\rm el}(t), \\
\frac{dE_{\rm ph1}(t)}{dt}&=I_{\rm ph1}(t).
\end{align}
$I_{\rm el}$ and $I_{\rm ph1}$ indicates the energy flow between subsystems, and $W_{\rm el}$ means the contribution of the external field.
Detailed expressions of $I_{\rm el}$, $I_{\rm ph1}$, and $W_{\rm el}$ are
 \begin{align}
I_{\rm el}(t)&=I^{\Sigma_{\rm el-ph1}}(t)+I^{\Sigma_{\rm el-ph2}}(t), \label{eq:Iel}\\
I_{\rm ph1}(t)&=I^{\Pi_{\rm el-ph1}}(t)+I^{\Pi_{\rm ph1-ph2}}(t), \label{eq:Iph1}\\
I^{\Sigma_{\rm el-ph1}}(t)&=-4{\rm Re}\left\{ \left[ \Sigma_{\rm el-ph1} \ast G \ast \Delta \right]^< (t,t) \right\}, \label{eq:Iph1el}\\
I^{\Sigma_{\rm el-ph2}}(t)&=-4{\rm Re}\left\{ \left[ \Sigma_{\rm el-ph2} \ast G \ast \Delta \right]^< (t,t) \right\}, \label{eq:Iph2el}\\
I^{\Pi_{\rm el-ph1}}(t)&={\rm Im}\left\{ \left. \frac{\partial}{\partial t} \left[D_1\ast \Pi_{\rm el-ph1} \right]^<(t,t')\right|_{t'=t}\right\}, \label{eq:Ielph1}\\
I^{\Pi_{\rm ph1-ph2}}(t)&={\rm Im}\left\{ \left. \frac{\partial}{\partial t} \left[D_1\ast \Pi_{\rm ph1-ph2} \right]^<(t,t')\right|_{t'=t}\right\}, \label{eq:Iph2ph1} \\
W_{\rm el}(t)&=
\frac{d\cos\phi(t)}{dt}\frac{1}{\cos\phi(t)}\left\{ -2i\left[ \Delta_1 \ast G \right]^{<} (t,t)\right\} \nonumber \\
&+\frac{d\sin\phi(t)}{dt}\frac{1}{\sin\phi(t)}\left\{ -2i\left[ \Delta_2 \ast G \right]^{<} (t,t)\right\}.
\end{align}
Here, $\Delta_1(t,t')=J^{*}\cos\phi(t)G(t,t')J^{*}\cos\phi(t')$ and $\Delta_2(t,t')=J^{*}\sin\phi(t)G(t,t')J^{*}\sin\phi(t')$.
$I^{\Sigma_{\rm el-ph1}}$ and $I^{\Pi_{\rm el-ph1}}$ represent the energy flow between the electron and phonon-1.
Note that magnitudes of $I^{\Sigma_{\rm el-ph1}}$ and $I^{\Pi_{\rm el-ph1}}$ are different, i.e., $I^{\Sigma_{\rm el-ph1}} \neq -I^{\Pi_{\rm el-ph1}}$, and this difference causes the time-evolution of the electron-phonon-1 interaction energy $E_{\rm el-ph1}$ (see the more detailed discussion in Appendix \ref{app:interaction}).
$I^{\Sigma_{\rm el-ph2}}$ ($I^{\Pi_{\rm ph1-ph2}}$) represents the energy flow between the electron (phonon-1) and the phonon-2.

In the following, we use $I^\Sigma$ to express either of $I^{\Sigma_{\rm el-ph1}}$ and $I^{\Sigma_{\rm el-ph2}}$. 
This is because the expressions of $I^{\Sigma_{\rm el-ph1}}$ and $I^{\Sigma_{\rm el-ph2}}$ are almost the same except for the electron self-energy used. 
Due to the similar reason,  we use $I^{\Pi}$ to express $I^{\Pi_{\rm el-ph1}}$ or $I^{\Pi_{\rm ph1-ph2}}$.

\subsubsection{Approximated energy flow equations}
In this section, we derive the energy flow equations of different levels of approximations for the electron-phonon coupled system.
The most simplified equations are based on the gradient and quasiparticle approximations, and they can be further simplified to the BE and the MTM using the quasiparticle approximation of the self-energy and the quasiequilibrium approximation (see Appendix \ref{app:GFtoBE}).

First, we focus on electrons and derive the equations with the following idea. 
At long enough times after the photo-excitation  ($t \gg  t_p$, $t_p$: pump time), the time evolution of the system becomes relatively mild, where the concept of the nonequilibrium distributions of the Green's functions and the self-energies becomes meaningful. 
Previous studies~\cite{PhysRevX.8.041009,FREERICKS2021147104} showed that, in this regime, the kinetic equation of electron-population $n_k(t)=\langle \hat{c}^{\dagger}_k(t)\hat{c}_k(t) \rangle=-iG_k^<(t,t)$ ($k$ is the index of eigenstates)
is ruled by the difference of the nonequilibrium distribution functions of the electron Green's function and the electron self-energy. 
This indicates that  the electron energy flow $I^{\Sigma}$ is also ruled by the difference of the nonequilibrium distribution functions because the kinetic equation of electron energy is expressed as $\frac{dE_{\rm kin}}{dt}=\frac{2}{N}\sum_k\epsilon_k \frac{dn_k}{dt}$ where $\epsilon_k$ is the eigenenergy.

The above idea is mathematically formulated using the Winger representation.
The Wigner representation of a two-time function $A(t,t')$ is defined as 
\begin{align}
A(\omega,t_a)=\int_{-\infty}^{\infty}A\left(t_a+\frac{t_r}{2},t_a-\frac{t_r}{2}\right)e^{i\omega t_r}dt_r, \label{eq:Wigner}
\end{align}
where $t_a=\frac{t+t'}{2}$ is the average time and $t_r=t-t'$ is the relative time.
In the Winger representation, the convolution of two-time functions are expressed with the Moyal product~\cite{Mahan2000,10.1143/PTP.123.581,kamenev_2023,PhysRevB.98.134312,PhysRevB.104.085108}.
We express $\frac{dn_k}{dt}$ with the Wigner transforms of the Green's function and the self-energy.
Keeping terms up to the first order with respect  to differential operators of time ($\partial_t$) and energy ($\partial_\omega$) (the gradient expansion) (see the detailed calculation in Refs.~\cite{Mahan2000,10.1143/PTP.123.581,https://doi.org/10.1002/prop.201600042}),
we obtain the approximated expression of the energy flow $I^{\Sigma}$ as
\begin{widetext}
\begin{align}
I^{\Sigma}(t) 
&\approx
-\frac{8}{N}\sum_k\epsilon_k
\int_{-\infty}^{\infty}\frac{d\omega}{2\pi}\left({\rm Im}G_k^R(\omega,t)\right)\left({\rm Im}\Sigma^R(\omega,t)\right)
\left(f^{G_k}(\omega,t)-f^\Sigma(\omega,t)\right) \nonumber \\
&+\frac{2}{N}\sum_k\epsilon_k\int_{-\infty}^{\infty}\frac{d\omega}{2\pi}
\left[
\left\{{\rm Re}\Sigma^R(\omega,t), {\rm Im}G_k^<(\omega,t)\right\}_p
 -\left\{{\rm Re}G_k^R(\omega,t), {\rm Im}\Sigma^<(\omega,t)\right\}_p
 \right]. \label{eq:Ielapprox_Gk}
\end{align}
\end{widetext}
$\left\{A,B\right\}_P\equiv\partial_\omega A\partial_t B-\partial_t A\partial_\omega B$ is the Poisson bracket.
Two nonequilibrium distribution functions $f^{G_k}$ and $f^\Sigma$ are defined as
\begin{align}
f^{G_k}(\omega,t)&\equiv\frac{{\rm Im}G^<_{k}(\omega,t)}{-2{\rm Im}G^R_{k}(\omega,t)}, \\
f^{\Sigma}(\omega,t)&\equiv \frac{{\rm Im}\Sigma^<(\omega,t)}{-2{\rm Im}\Sigma^R(\omega,t)}.
\end{align}
In equilibrium, these two functions are equal to the Fermi-Dirac distribution function (the fluctuation dissipation theorem).
In the gradient approximation ignoring the Poisson bracket terms, one can understand that the energy flow is ruled by the difference of the nonequilibrium distribution functions of the Green's function and the self-energy~\cite{PhysRevX.8.041009,FREERICKS2021147104}.

We consider the further approximation of $I^\Sigma$ partially using the quasiparticle approximation.
We introduce the nonequilibrium distribution function of the local Green's function $G(t,t')=\frac{1}{N}\sum_kG_k(t,t')$ as 
\begin{align}
f^G(\omega,t)
&\equiv \frac{{\rm Im}G^<(\omega,t)}{-2{\rm Im}G^R(\omega,t)} \nonumber \\
&=\frac{1}{N}\frac{\sum_k\left\{-\frac{1}{\pi}{\rm Im}G_k^R(\omega,t)\right\}f^{G_k}(\omega,t)}{\frac{1}{N}\sum_k-\frac{1}{\pi}{\rm Im}G_k^R(\omega,t)}.
\end{align}
If the interaction is weak and the quasiparticle approximation is valid, i.e., $-\frac{1}{\pi}{\rm Im}G_{k}^R(\omega,t) \to \delta(\omega-\epsilon_k)$~\cite{10.1143/PTP.123.581},
$f^G$ can be approximated as
\begin{align}
f^G(\omega,t)
&\approx
\frac{1}{N}\frac{\sum_k\left\{\delta(\omega-\epsilon_k)\right\}f^{G_k}(\omega,t)}{\frac{1}{N}\sum_k\delta(\omega-\epsilon_k)}  \nonumber \\
&=
\frac{1}{\rho(\omega)}\frac{1}{N}\sum_k\delta(\omega-\epsilon_k)f^{G_k}(\omega,t). \label{eq:fGapprox}\\ 
&\left(=\frac{1}{\rho(\omega)}\frac{1}{N}\sum_k\delta(\omega-\epsilon_k)f^{G_k}(\epsilon_k,t)\right) \nonumber 
\end{align}
Here, $\rho(\omega)=\sum_k\delta(\omega-\epsilon_k)/N$ is the density of states of free electrons.
The similar treatment has been used in the study using the BE to derive the distribution function depend on $\omega$~\cite{PhysRevB.78.174514}.
Using above approximations, $I^\Sigma$ is expressed as
\begin{align}
I^{\Sigma}(t)
&\approx
4\int_{-\infty}^{\infty}\omega\left({\rm Im}\Sigma^R(\omega,t)\right)\rho(\omega)
\left\{f^G(\omega,t)-f^\Sigma(\omega,t)\right\}d\omega. \label{eq:Ielapprox}
\end{align}
This equation means that the time-evolution of the electron kinetic energy is ruled by the difference of the nonequilibrium distribution functions of 
the local Green's function and the self-energy in the weak coupling regime.
Note that the above argument is applied to each of $\Sigma=\Sigma_{\rm el-ph1}$ and $\Sigma=\Sigma_{\rm el-ph2}$.

As for the energy flow $I^{\Pi}$, assuming relatively slow dynamics of the system, we can also derive the approximated expression involving the difference of two nonequilibrium bosonic distribution functions for the phonon Green's function and the phonon self-energy. 
To derive the expression, we use the two types of the time-derivative of phonon-1 Green's function $D_1$: $D_{d_1}(t,t')\equiv \partial_tD_1(t,t')/\omega_1$ and $D_{d_2}(t,t')\equiv \partial_{t'}D_1(t,t')/\omega_1$~\cite{PhysRevB.91.045128}.
Using these Green's functions, the energy flow $I^{\Pi}$ can be expressed as
\begin{widetext}
\begin{align}
I^{\Pi}(t)
&={\rm Im}\left\{ \omega_1\left[D_{d_1}\ast \Pi\right]^<(t,t)\right\} \nonumber \\
&= \frac{-i \omega_1}{2}\int_{-\infty}^{\infty}
\left\{
D_{d_1}^R(t,\bar{t})\Pi^<(\bar{t},t)+D_{d_1}^<(t,\bar{t})\Pi^A(\bar{t},t)
+\Pi^R(t,\bar{t})D_{d_2}^<(\bar{t},t)+\Pi^<(t,\bar{t})D_{d_2}^A(\bar{t},t)
\right\}d\bar{t}.
\end{align}
Assuming $t\gg t_p$ (assuming slow dynamics of the Green's function and the self-energy) as in the case of electrons~\cite{e18050180,https://doi.org/10.1002/prop.201600042,PhysRevX.8.041009}, 
we apply the gradient expansion using the Wigner transformation and obtain 
\begin{align}
I^{\Pi}(t)
&\approx
\omega_1\int_{-\infty}^{\infty}\frac{d\omega}{2\pi}
\left\{
\left({\rm Re}D_{d_1}^R(\omega,t)\right)\left({\rm Im}\Pi^<(\omega,t)\right)
-\left({\rm Im}\Pi^R(\omega,t)\right)\left({\rm Re}D_{d_1}^<(\omega,t)\right) 
+\left({\rm Re}\Pi^R(\omega,t)\right)\left({\rm Im}D_{d_1}^<(\omega,t)\right)
\right\} \nonumber \\
&+\omega_1\int_{-\infty}^{\infty}\frac{d\omega}{4\pi}
\left[
\left\{{\rm Re}D_{d_1}^<(\omega,t),{\rm Re}\Pi^R(\omega,t)\right\}_P 
-\left\{{\rm Im}D_{d_1}^R(\omega,t),{\rm Im}\Pi^<(\omega,t)\right\}_P
+\left\{{\rm Im}D_{d_1}^<(\omega,t),{\rm Im}\Pi^R(\omega,t)\right\}_P
\right]. \label{eq:Iph1_Dd1}
\end{align}
\end{widetext}
The Wigner transforms of $D_{d_1}^<$ and $D_{d_1}^R$ can be expressed as 
\begin{align}
D^R_{d_1}(\omega,t_a)=\frac{1}{2\omega_1}\partial_{t_a}D_1^R(\omega,t_a)-i\frac{\omega}{\omega_1}D_1^R(\omega,t_a), \\
D^<_{d_1}(\omega,t_a)=\frac{1}{2\omega_1}\partial_{t_a}D_1^<(\omega,t_a)-i\frac{\omega}{\omega_1}D_1^<(\omega,t_a).
\end{align}
When the dynamics of the system is slow enough, we can approximately ignore terms with  $\partial_{t_a}$,
\begin{align}
D_{d_1}^R(\omega,t_{\rm ave})\approx-i\frac{\omega}{\omega_1}D_1^R(\omega,t_{\rm ave}), \label{eq:Dd1approx1}\\
D_{d_1}^<(\omega,t_{\rm ave})\approx-i\frac{\omega}{\omega_1}D_1^<(\omega,t_{\rm ave}). \label{eq:Dd1approx2}
\end{align}
Within this approximation, $D_{d_1}^<(\omega,t_{\rm ave})$ becomes real, i.e., ${\rm Im}D_{d_1}^<(\omega,t_{\rm ave}) \approx 0$, because $D_1^<(\omega,t_{\rm ave})$ is purely imaginary.
If we additionally ignore the Poisson bracket terms,
we obtain the expression of $I^\Pi$ similar to $I^\Sigma$ (Eq.~\eqref{eq:Ielapprox}),
\begin{align}
&I^{\Pi}(t)\approx \nonumber \\
&\int_{-\infty}^{\infty}\omega\left({\rm Im}\Pi^R(\omega,t)\right)\left(-{\rm Im}D_1^R(\omega,t)/\pi\right)\left(f^{D_1}(\omega,t)-f^{\Pi}(\omega,t)\right) d\omega. \label{eq:Iph1approx}
\end{align}
Here, $f^{D_1}$ and $f^{\Pi}$ are defined as
\begin{align}
f^{D_1}(\omega,t)&\equiv\frac{{\rm Im}D_1^<(\omega,t)}{2{\rm Im}D_1^R(\omega,t)}, \\
f^{\Pi}(\omega,t)&\equiv\frac{{\rm Im}\Pi^<(\omega,t)}{2{\rm Im}\Pi^R(\omega,t)}.
\end{align}
These two functions are equal to the Bose-Einstein distribution function in equilibrium.
Note that the above argument is applied to each of $\Pi=\Pi_{\rm el-ph1}$ and $\Pi=\Pi_{\rm ph1-ph2}$.

\subsection{Discussion}
In the previous sections, we derived the expressions of the energy flow $I$ at different levels of approximations assuming the slow dynamics of the system.
In particular, Eq.~\eqref{eq:Ielapprox} is based on the quasiparticle approximation for the Green's function and the gradient approximation. 
In addition, Eq.~\eqref{eq:Iph1approx} is based on the gradient approximations.
In which situation these approximations are justified in practice is one of the main subjects to be discussed in the following.

In addition, Eqs.~\eqref{eq:Ielapprox} and \eqref{eq:Iph1approx} can be further simplified to the BE or the MTM, see Appendix \ref{app:GFtoBE}.
Namely, with the quasiparticle approximations to the self-energies leads to the BE.
Furthermore, additional quasiequilibrium approximation, where we assume that the nonequilibrium distribution functions of electrons and phonons can be approximated with the Fermi-Dirac distribution or the Bose-Einstein distribution with time-dependent temperatures $T_{\rm el}(t)$ and $T_{\rm ph1}(t)$, leads to the MTM. 
In this case, the nonequilibrium distribution $f^{\Pi_{\rm el-ph1}}$ matches to $1/(\exp(\omega/T_{\rm el}(t))-1)$ with the time-dependent electron temperature $T_{\rm el}(t)$ ($f^{\Pi_{\rm ph1-ph2}}$ exactly equals to the Bose-Einstein distribution with the bath-phonon temperature).
This makes the interpretation of the energy flow $I^{\Pi}$ using the MTM possible.
In the following, we also discuss in which situation the interpretation using the MTM is reasonable in practice.

For the discussion, we evaluate the effective temperatures from the nonequilibrium distribution functions of the Green's function and the self-energy, both of which we express as $A$ here.
First, we set the fitting range $\mathbb{F}$ and calculate the value $T^A_i$ for each $\omega_i \in \mathbb{F}$ as
\begin{align}
T^A_i(t)=\frac{\omega_i}{{\rm log}\left\{1/f^A(\omega_i,t)\mp 1\right\}} \ (-/+:\ {\rm Fermion/Boson}).
\label{eq:eacheffectiveT}
\end{align}
We calculate the average $T^A_{\rm AVE}$ and the standard deviation $T^A_{\rm SD}$ from datasets $\left\{T^A_i(t)\right\}_{\mathbb{F}}$.
Then, we define the nonequilibrium effective temperature $T^A$ as
\begin{align}
T^A(t)\equiv T^A_{\rm AVE}(t)\pm T^A_{\rm SD}(t).
\label{eq:effectiveT}
\end{align}

\section{Results}\label{sec:results}
\begin{figure}
\centering
\includegraphics[scale=0.35]{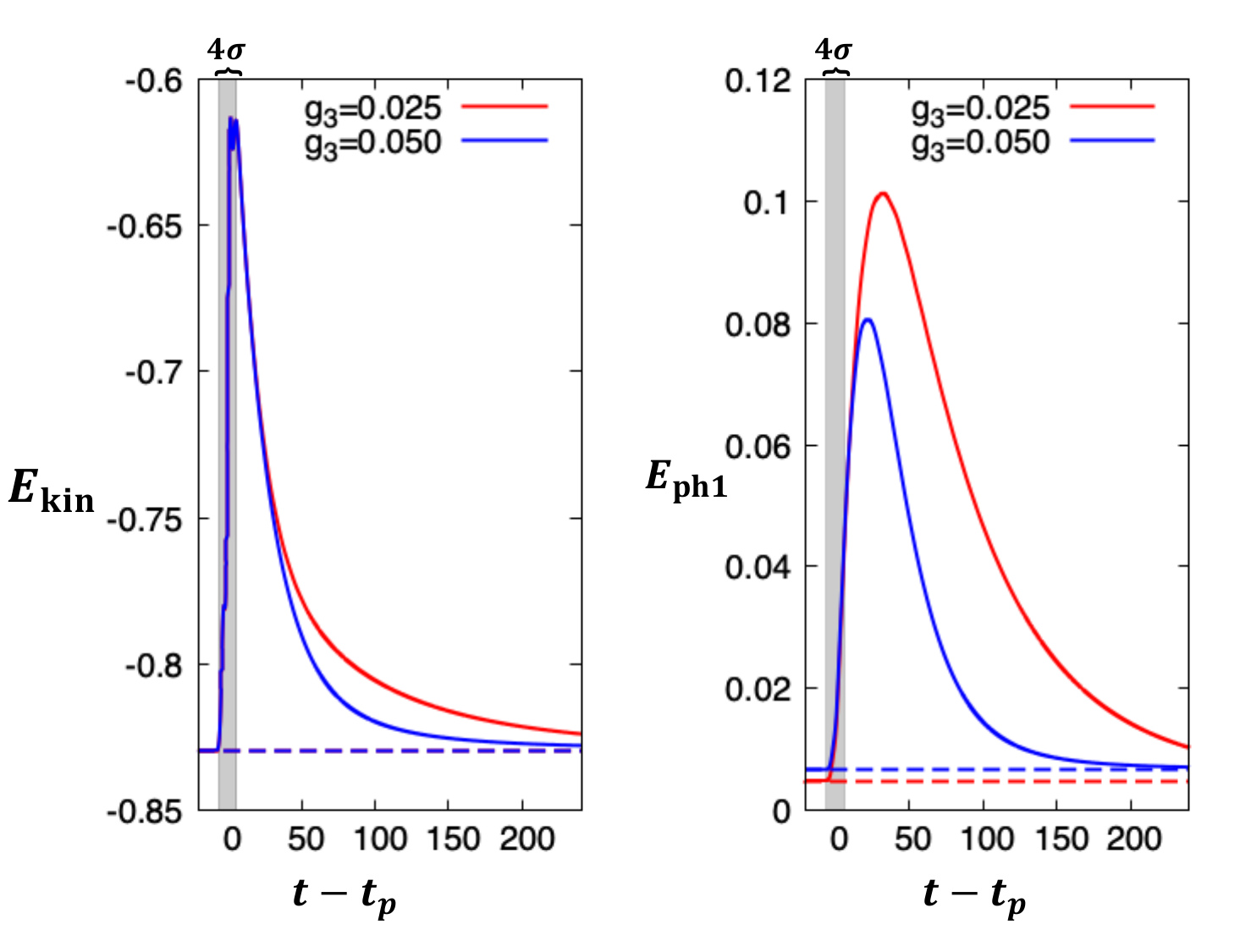}
\caption{(Color online) Time-evolution of the electron kinetic energy $E_{\rm kin}$ and the phonon-1 energy $E_{\rm ph1}$ for $g_1=g_2=0.2$, $\omega_1=0.4$, and $\beta=20$.
For the bath parameters, we set $\omega_D=0.4$ and $\gamma=0.1$.
For the electric field, we set $E_{\rm el}=0.4$, $\Omega_{\rm el}=0.4$, and $\sigma=3$.}
\label{energy}
\end{figure}
In this paper, unless otherwise noted, we set the equilibrium temperature $\beta=20$ of the initial state and 
the phonon-1 frequency $\omega_1=0.4$. 
Note that the bath-phonon temperature is equal to the equilibrium temperature.
For the heat-bath phonon, we use $\omega_D=0.4$ and $\gamma=0.1$.
For the electric field, we use the vector potential $A(t)$ expressed as 
\begin{align}
A(t)&=\frac{E_{\rm el}}{\Omega_{\rm el}}\sin(\Omega_{\rm el}(t-t_p))\exp\left\{-\frac{(t-t_p)^2}{2\sigma^2}\right\}.
\end{align}
We set $E_{\rm el}=0.4$, $\Omega_{\rm el}=0.4$, $\sigma=3$.

In equilibrium, the coupling strength between electrons and two phonon modes is quantified by the parameter $\lambda$,
\begin{align}
\lambda&=\lambda_1+\lambda_2, \\
\lambda_1&=\left. -\frac{\partial}{\partial \omega}{\rm Re}\Sigma_{\rm el-ph1}^R(\omega)\right|_{\omega=0} , \\
\lambda_2&=\left. -\frac{\partial}{\partial \omega}{\rm Re}\Sigma_{\rm el-ph2}^R(\omega)\right|_{\omega=0}.
\end{align}
The effective mass of electron is estimated as $m^*/m=(1+\lambda)$.
In this work, we set the parameter $\lambda <1$ so that the Migdal approximation is valid~\cite{PhysRevX.3.041033,PhysRevB.98.245110}.

In the following sections, we study the energy flows in different electron-phonon coupling regimes.
In particular, we focus on the backward energy flow where the direction of the energy flow between electrons and hot-phonons (phono-1) is reversed in time.
This phenomenon is one of examples of complex energy flows in systems with electrons coupled to multiple phonon modes.
By comparing the full energy flows expressed as  Eqs.~\eqref{eq:Iph1el}-\eqref{eq:Iph2ph1} with those of approximated expressions, 
we discuss the validity of approximations and the physcial picture of the energy flows that these approximated expressions provide.
In particular, the expressions of  Eqs.~\eqref{eq:Ielapprox} and \eqref{eq:Iph1approx} tell that the energy flows can be understood from the nonequilibrium distribution functions. 
Such picture is closely related to the picture provided by the MTM and we also discuss its validity.

\subsection{Backward energy flow in weak electron-phonon coupling regime}\label{results:backflow}
In this subsection, we consider systems in the weak electron-phonon coupling regime.
Here, we mainly use $g_1=g_2=0.2$ and change the phonon-phonon coupling $g_3$.
In these parameters, the coupling strength is $\lambda\approx0.12 \ (\lambda_1\approx 0.068,\ \lambda_2\approx 0.05)$ for $g_3=0.025$ and 
$\lambda\approx0.12 \ (\lambda_1\approx 0.072,\ \lambda_2\approx 0.049)$ for $g_3=0.05$.

First, we show the time-evolution of the electron kinetic energy $E_{\rm kin}$ and the phonon-1 energy $E_{\rm ph1}$, which are introduced in Eqs.~\eqref{eq:Ekin} and \eqref{eq:Eph1},  in Fig.~\ref{energy}.
In equilibrium, $E_{\rm kin}$ and $E_{\rm ph1}$ are quantitatively almost the same for $g_3=0.025$ and $g_3=0.05$.
The electron kinetic energy $E_{\rm kin}$ increases during the photo-excitation, and then decreases toward its equilibrium value after the excitation.
Just after the photo-excitation, the decrease rate of $E_{\rm kin}$ are almost independent of $g_3$. 
However, in the later stage, the stronger the phonon-phonon coupling $g_3$ is, the faster $E_{\rm kin}$ approaches its equilibrium value.
The phonon-1 energy $E_{\rm ph1}$ increases up to a certain maximum value during and after the photo-excitation of electrons.
In the later stage, $E_{\rm ph1}$ decreases toward its equilibrium value as $E_{\rm kin}$.
The corresponding time-evolution of the nonequilibrium spectral functions of the Green's functions and the self-energies is shown in  Appendix~\ref{app:spectrum}.

\begin{figure*}
\centering
\includegraphics[scale=0.4]{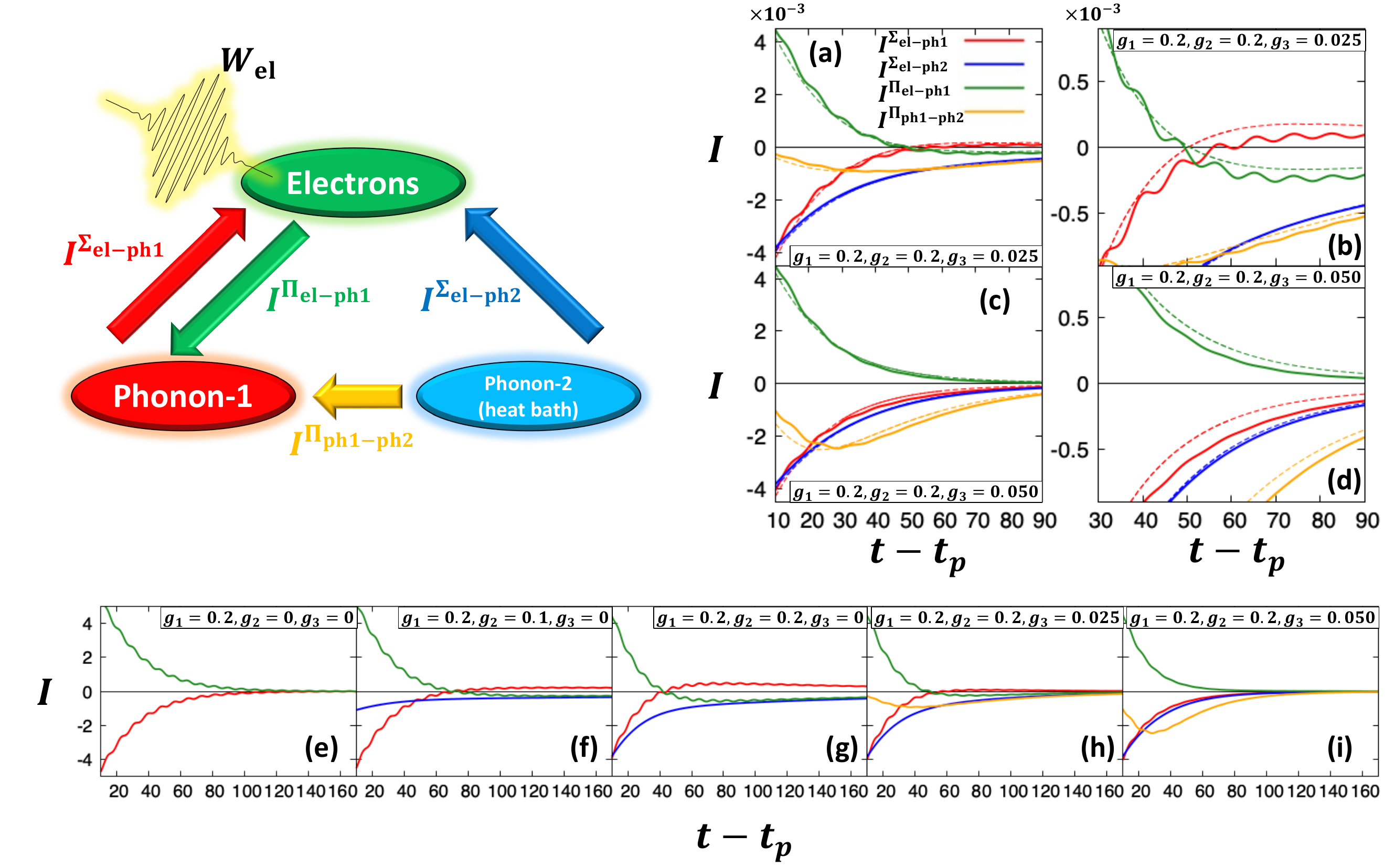}
\caption{
(Color online) Left figure shows the schematic picture of energy flows. 
(a) Energy flows for $g_1=0.2$, $g_2=0.2$, and $g_3=0.025$. (b) is the enlarged view of (a).
(c) Energy flows for $g_1=0.2$, $g_2=0.2$, and $g_3=0.05$. (d) is the enlarged view of (c).
We set $\omega_1=0.4$, and $\beta=20$.
For the bath parameters, we set $\omega_D=0.4$ and $\gamma=0.1$.
For the electric field, we set $E_{\rm el}=0.4$, $\Omega_{\rm el}=0.4$, and $\sigma=3$.
(e)-(i) The comparison of energy flows in the normal Holstein model (e), those in the two-phonon coupling Holstein model (f)(g), and those in the present model (h)(i) for the same excitation condition.
Solid lines represent $I^\Sigma$ and $I^\Pi$ evaluated from Eqs.~(\ref{eq:Iph1el})-(\ref{eq:Iph2ph1}), and dashed lines represent approximated values evaluated from Eqs.~(\ref{eq:Ielapprox}) and (\ref{eq:Iph1approx}).
}
\label{energyflow}
\end{figure*}
In order to examine the detailed energy transfer between subsystems, we evaluate the energy flows.
The full energy flows evaluated from  Eqs.~\eqref{eq:Iph1el}-\eqref{eq:Iph2ph1} are shown with solid lines in Fig.~\ref{energyflow}.
First, we consider $g_3=0.025$ (Figs.~\ref{energyflow} (a) and (b)).
Characteristic oscillatory behavior is observed  in $I^{\Sigma_{\rm el-ph1}}$, $I^{\Pi_{\rm el-ph1}}$, and $I^{\Pi_{\rm ph1-ph2}}$ while such behavior is absent in $I^{\Sigma_{\rm el-ph2}}$.
The frequency of these oscillations ($\omega_{osc}\approx 0.74$) is about twice of the phonon-1 frequency renormalized by the electron-phonon or phonon-phonon couplings ($\omega_1^r \approx 0.365$).
Similar oscillatory behavior has been reported in several theoretical studies~\cite{PhysRevB.91.045128,PhysRevB.98.245110},
which is attributed to the dynamics of phonon-1 (see Appendix~\ref{app:oscillation}).
In the initial relaxation process just after the photo-excitation, both $I^{\Sigma_{\rm el-ph1}}$ and $I^{\Sigma_{\rm el-ph2}}$ are negative.
This means that, in the initial process, the electron energy is transferred to the phonon-1 and the phonon-2.
Inversely, the phonon-1 obtains the energy from electrons, which makes $I^{\Pi_{\rm el-ph1}}$ positive.
Simultaneously, the phonon-1 energy flows to the heat-bath, and $I^{\Pi_{\rm ph1-ph2}}$ becomes negative.
$E_{\rm ph1}$ takes the maximum value around  $t-t_p \approx 32$ due to the competition of $I^{\Pi_{\rm el-ph1}}$ and $I^{\Pi_{\rm ph1-ph2}}$~\cite{PhysRevB.97.054310}.
In the later stage, the energy flow $I^{\Sigma_{\rm el-ph1}}$ ($I^{\Pi_{\rm el-ph1}}$) changes from negative (positive) to positive (negative) at $t-t_p\approx55$ ($t-t_p\approx 49$), see Fig.~\ref{energyflow} (b).
These mean that the direction of the energy flow between  electrons and phonon-1 is reversed.
This backward energy flow has been reported in previous studies using the BE~\cite{PhysRevB.97.054310,ShotaOno2019,PhysRevB.101.100302,PhysRevB.103.125412} and the MTM~\cite{PhysRevX.6.021003,PhysRevB.96.174439,PhysRevB.98.134309,PhysRevB.101.035128,PhysRevLett.124.077001,PhysRevB.101.100302,SciPostPhys.12.5.173}.

On the other hand, for $g_3=0.05$, unlike for $g_3=0.025$, both $I^{\Sigma_{\rm el-ph1}}$ and $I^{\Sigma_{\rm el-ph2}}$ are always negative, and the backward energy flow is not observed, see Figs.~\ref{energyflow}(c) and (d).
To clarify the condition when the backward energy flow happens, in Figs.~\ref{energyflow}(e)-(i), we compared the cases of the normal Holstein model ($g_1=0.2, \ g_2=g_3=0$), the Holstein models with two-phonon modes ($g_1=0.2,\ g_2\neq 0, \ g_3=0$), and the present model  ($g_1=0.2,\ g_2=0.2, \ g_3\neq0$) for the same excitation condition.
In the normal Holstein model ($g_1=0.2, \ g_2=g_3=0$), we cannot see the backward energy flow, see Fig.~\ref{energyflow}(e).
On the other hand, for the cases with $g_2\neq 0, \ g_3=0$, the backward energy flow appears, see Figs.~\ref{energyflow}(f) and (g).
With increasing $g_3$, the dissipation of $E_{\rm ph1}$ to the phonon-2 (heat-bath) suppresses the backward energy flow, see Figs.~\ref{energyflow}(h) and (i).
These comparison confirms that the backward energy flow indeed originates from the existence of the phonon-2 and the excess energy stored in the phonon-1.

\begin{figure*}
\centering
\includegraphics[scale=0.65]{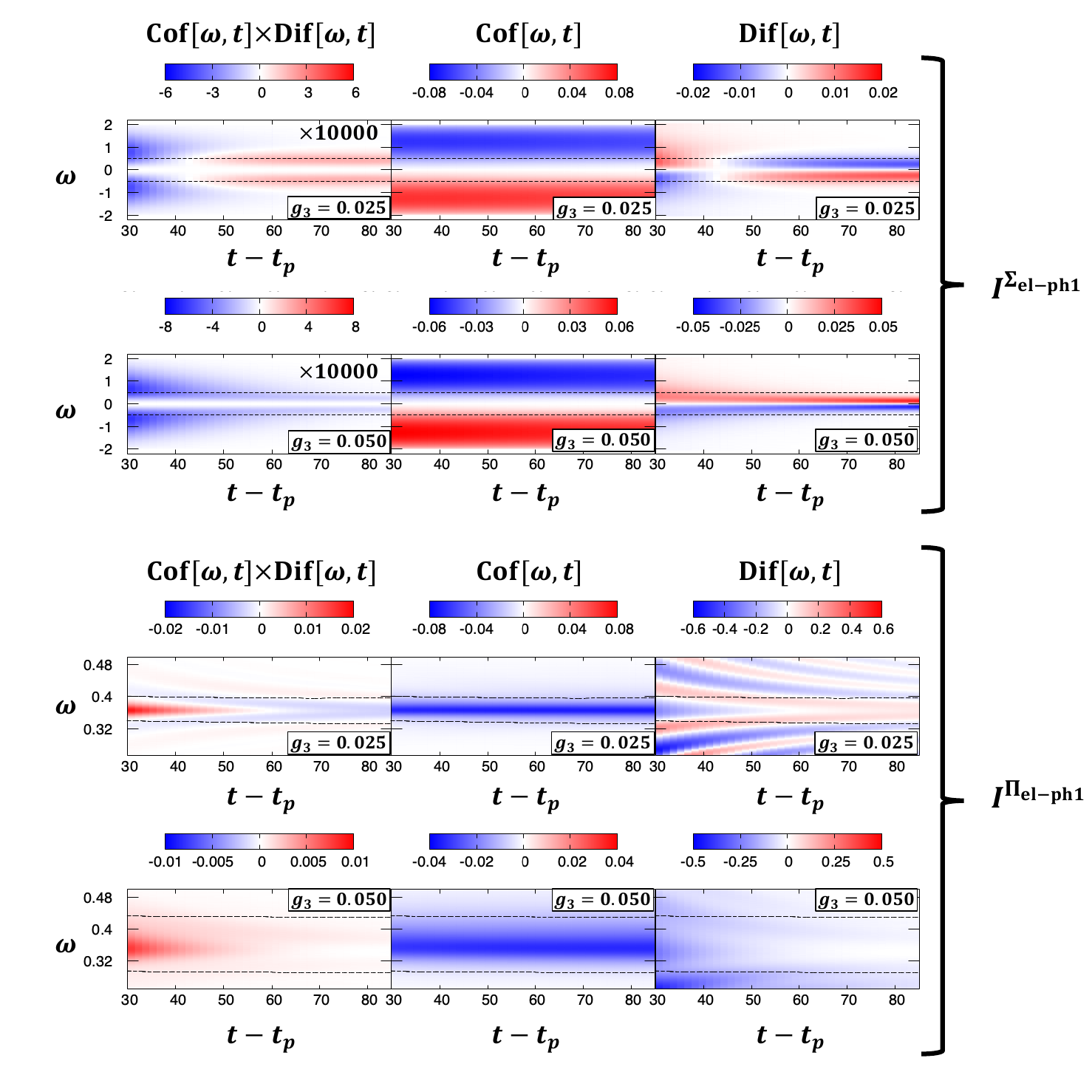}
\caption{
(Color online) Each parts of integral function of $I^{\Sigma_{\rm el-ph1}}$ and $I^{\Pi_{\rm el-ph1}}$ expressed as Eqs.~(\ref{eq:eachparts}) and (\ref{eq:eachparts2}) for $g_1=0.2$, $g_2=0.2$.
We set $\omega_1=0.4$, and $\beta=20$.
For the bath parameters, we set $\omega_D=0.4$ and $\gamma=0.1$.
For the electric field, we set $E_{\rm el}=0.4$, $\Omega_{\rm el}=0.4$, and $\sigma=3$.
The energy ($\omega$) region between two black dashed lines is the fitting range $\mathbb{F}$ for the effective temperatures.
}
\label{energyflowanalysis}
\end{figure*}

In order to obtain the microscopic picture of energy flows, we compare the approximated energy flows calculated from Eqs.~\eqref{eq:Ielapprox} and \eqref{eq:Iph1approx}, which are ruled by the nonequilibrium distribution functions,
with the full energy flows, see Fig.~\ref{energyflow}.
We can see that the approximated energy flows (dashed lines in Fig.~\ref{energyflow}) well reproduce the full energy flows (solid lines in Fig.~\ref{energyflow}) both for $g_3=0.025$ and $g_3=0.05$.
These results suggest that the gradient approximation and the quasiparticle approximation used for Eqs.~\eqref{eq:Ielapprox} and \eqref{eq:Iph1approx} are justified and these approximated expressions include the essence of the backward energy flow.
We note that the magnitudes of the approximated energy flows $I^{\Sigma_{\rm el-ph1}}$ (red dashed line) and $I^{\Pi_{\rm el-ph1}}$ (green dashed line) are almost the same, i.e., $I^{\Sigma_{\rm el-ph1}}\simeq-I^{\Pi_{\rm el-ph1}}$.
This indicates that the approximated energy flows approach to those of the BE, where $I^{\Sigma_{\rm el-ph1}}=-I^{\Pi_{\rm el-ph1}}$ are exactly fulfilled~\cite{PhysRevB.97.054310}.
Note again that, in the NEGF, magnitudes of $I^{\Sigma_{\rm el-ph1}}$ and $I^{\Pi_{\rm el-ph1}}$ are different in general, and this difference causes the time-evolution of the interaction energy between electrons and phonon-1 (see Appendix~\ref{app:interaction}).

\begin{figure*}
\centering
\includegraphics[scale=1.2]{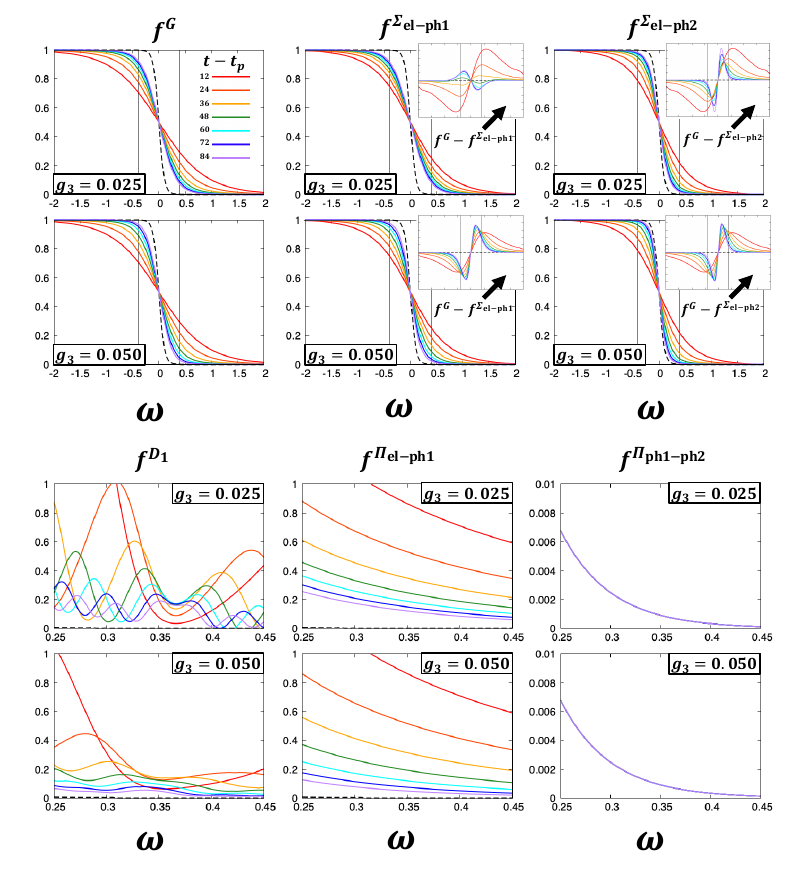}
\caption{
(Color online) Time-evolution of the nonequilibrium distribution functions of the Green's functions and the self-energies for $g_1=0.2$, $g_2=0.2$.
We set $\omega_1=0.4$, and $\beta=20$.
For the bath parameters, we set $\omega_D=0.4$ and $\gamma=0.1$.
For the electric field, we set $E_{\rm el}=0.4$, $\Omega_{\rm el}=0.4$, and $\sigma=3$. 
Small inset figures represent the difference between $f^G$ and $f^{\Sigma}$.
Black dashed lines show equilibrium values (fermion: $1/({\rm exp}(\beta\omega)+1)$, boson: $1/({\rm exp}(\beta\omega)-1))$.
Black horizontal lines in $f^G$, $f^{\Sigma_{\rm el-ph1}}$, and $f^{\Sigma_{\rm el-ph2}}$ represent the phonon window, i.e., $\omega \in [-\omega_1,\omega_1]$ ($\omega_1=\omega_D$). 
}
\label{noneqdistribution}
\end{figure*}

\begin{figure}
\centering
\includegraphics[scale=0.4]{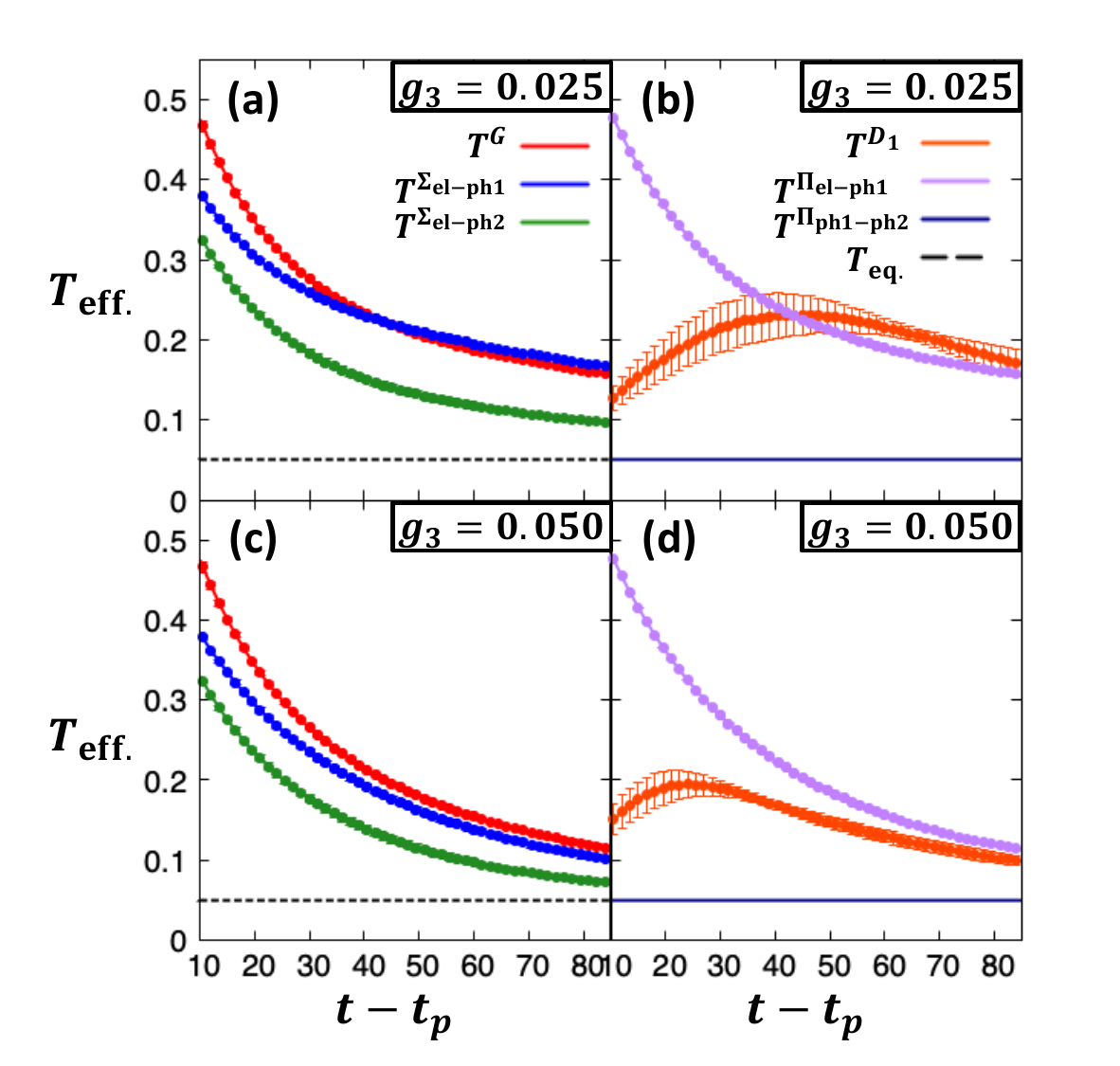}
\caption{
(Color online) Nonequilibrium effective temperatures of the Green's functions and the self-energies expressed as Eq.~\eqref{eq:effectiveT} for $g_1=0.2$, $g_2=0.2$.
We set $\omega_1=0.4$, and $\beta=20$.
For the bath parameters, we set $\omega_D=0.4$ and $\gamma=0.1$.
For the electric field, we set $E_{\rm el}=0.4$, $\Omega_{\rm el}=0.4$, and $\sigma=3$.
(a) Electron effective temperatures for $g_3=0.025$. (b) Phonon-1 effective temperatures for $g_3=0.025$.
(c) Electron effective temperatures for $g_3=0.05$. (d) Phonon-1 effective temperatures for $g_3=0.05$.
Note that we plot standard deviations of $T^G$, $T^{\Sigma_{\rm el-ph1}}$, $T^{\Sigma_{\rm el-ph2}}$, and $T^{\Pi_{\rm el-ph1}}$, but these values are too small to see.
}
\label{effectiveT}
\end{figure}

Because the approximated energy flows qualitatively reproduce the full energy flows, the microscopic origin of the energy flow can be explained using the approximated expression.
In this study, we focus on the backward energy flow specifically.
Note that on the MTM the backward energy flow can be explained by the reversal of the electron temperature and the hot-phonon temperature.
On the other hand, the expressions with less approximations Eqs.~\eqref{eq:Ielapprox} and \eqref{eq:Iph1approx} tell that the nonequilibrium distribution functions of the Green's function and the self-energy are more general quantities to understand the energy flows~\cite{PhysRevX.8.041009,FREERICKS2021147104}. 
Thus, we examine the approximated $I^{\Sigma_{\rm el-ph1}}$ and $I^{\Pi_{\rm el-ph1}}$ specifically.
The integral functions of Eqs.~(\ref{eq:Ielapprox}) and (\ref{eq:Iph1approx}) can be split into the coefficient part (${\rm Cof}[\omega,t]$) and the difference part of the nonequilibrium distribution functions (${\rm Dif}[\omega,t]$) as 
\begin{align}
I^{\Sigma_{\rm el-ph1}}:
\begin{cases}
{\rm Cof}[\omega,t]=4\omega\left({\rm Im}\Sigma_{\rm el-ph1}^R(\omega,t)\right)\rho(\omega), \\
{\rm Dif}[\omega,t]=f^G(\omega,t)-f^{\Sigma_{\rm el-ph1}}(\omega,t), 
\end{cases} \label{eq:eachparts}
\end{align}
\begin{align}
I^{\Pi_{\rm el-ph1}}:
\begin{cases}
{\rm Cof}[\omega,t]=\omega\left({\rm Im}\Pi_{\rm el-ph1}^R(\omega,t)\right)\left(-{\rm Im}D_1^R(\omega,t)/\pi\right), \\
{\rm Dif}[\omega,t]=f^{D_1}(\omega,t)-f^{\Pi_{\rm el-ph1}} (\omega,t).
\end{cases} \label{eq:eachparts2}
\end{align}
In Fig. \ref{energyflowanalysis}, we show ${\rm Cof}[\omega,t]$ and ${\rm Dif}[\omega,t]$ for $I^{\Sigma_{\rm el-ph1}}$ and $I^{\Pi_{\rm el-ph1}}$. 
For both $g_3=0.025$ and $g_3=0.05$, the coefficient terms do not change the sign in time.
Namely, $I^{\Sigma_{\rm el-ph1}}$ is negative (positive) for $\omega>0$ ($\omega<0$) and $I^{\Pi_{\rm el-ph1}}$ is negative for $\omega>0$.
Therefore, the energy flow is mainly determined from the sign of ${\rm Dif}[\omega,t]$.
For $g_3=0.025$, in the initial relaxation process, ${\rm Dif}[\omega,t]$ for $I^{\Sigma_{\rm el-ph1}}$ is positive for $\omega>0$.
However, in the later stage, ${\rm Dif}[\omega,t]$ becomes negative.
This causes the sign change of $I^{\Sigma_{\rm el-ph1}}$ (the backward energy flow).
Similar sign-change behavior can be seen in the phonon-1 side ($I^{\Pi_{\rm el-ph1}}$).
The coefficient term of $I^{\Pi_{\rm el-ph1}}$ has the large weight around the renormalized frequency of the phonon-1 $\omega_1^r(t)$,  which is the peak position of the phonon spectrum $-{\rm Im}D_1^R(\omega,t)/\pi$ (see the phonon-1 spectrum in Appendix~\ref{app:spectrum}).
Therefore, we can estimate the sign of $I^{\Pi_{\rm el-ph1}}$ from the sign of ${\rm Dif}[\omega,t]$ around $\omega_1^r(t)$.
The difference part ${\rm Dif}[\omega,t]$ involves the positive and negative regions, but, in the vicinity of $\omega_1^r(t)$, the distribution of ${\rm Dif}[\omega,t]$ changes from negative to positive in time.
This causes the change of the sign of $I^{\Pi_{\rm el-ph1}}$ (the backward energy flow).
For $g_3=0.05$, neither  ${\rm Cof}[\omega,t]$ nor ${\rm Dif}[\omega,t]$ do not change the sign both for electrons and the phonon-1,
which explain the absence of the backward energy flow. 
From these analyses, in the weak electron-phonon coupling regime,  we conclude that the microscopic origin of the backward energy flow  can be understood as the reversal of the magnitude relation of $\left\{f^G, f^{\Sigma_{\rm el-ph1}}\right\}$ and $\left\{f^{D_1}, f^{\Pi_{\rm el-ph1}}\right\}$.

Now we discuss whether the above picture based on the distribution functions can be further reduced to the picture based on the temperatures.
In Fig.~\ref{noneqdistribution}, we show the time-evolution of the nonequilibirum distribution functions $f^G$, $f^{\Sigma}$, $f^{D_1}$, and $f^{\Pi}$.
We can estimate the effective temperatures from these distributions functions using Eq.~\eqref{eq:effectiveT}, which we show in Fig.~\ref{effectiveT}.
For the estimation, we use the fitting range $\mathbb{F}=[-0.5,0.5]$ for $f^G$ and $f^{\Sigma}$, and $\mathbb{F}=\left\{\omega \mid -{\rm Im}D_1^R(\omega,t)/\pi>0.2\times -{\rm Im}D_1^R(\omega_1^r(t),t)/\pi\right\}$ for $f^{D_1}$ and $f^{\Pi}$ (see black dotted lines in Fig.~\ref{energyflowanalysis}).
Note that the peak structure of $-{\rm Im}D_1^R(\omega,t)/\pi$ for $g_3=0.05$ is suppressed compared to that for $g_3=0.025$, so the fitting range for $g_3=0.05$ is larger than that for $g_3=0.025$.
In these fitting ranges, the inversion of the sign of ${\rm Dif}[\omega,t]$ can be clearly seen. 
One can see that $f^G$, $f^{\Sigma_{\rm el-ph1}}$, and $f^{\Sigma_{\rm el-ph2}}$ have the Fermi-Dirac distribution-like structure and $f^{\Pi_{\rm el-ph1}}$ has the Bose-Einstein distribution-like structure (see Fig.~\ref{noneqdistribution}).
Therefore, the standard derivations of effective temperatures for $f^G$, $f^{\Sigma_{\rm el-ph1}}$, $f^{\Sigma_{\rm el-ph2}}$, and $f^{\Pi_{\rm el-ph1}}$ are very small (these are too small to see in Fig.~\ref{effectiveT}).
Thus the effective temperature is well defined and the quasiequilbirum approximation is justified for these distributions.
On the other hand, $f^{D_1}$ strongly deviates from the Bose-Einsteion distribution so that the standard deviation of the estimated effective temperature $T^{D_1}$ is large compared to the rest.
In other words, the effective temperature is not a fully well-defined concept for this distribution function.
Still, we can try to argue the energy flow from the evolution of these effective temperatures. 

For $g_3=0.025$, for the electron part, we have $T^G>T^{\Sigma_{\rm el-ph1}}$ in the initial relaxation process, while, in the later stage, we have $T^G<T^{\Sigma_{\rm el-ph1}}$, see Fig.~\ref{effectiveT}(a).
Similarly, for the phonon-1 part,  the magnitude relation of $T^{D_1}$ and $T^{\Pi_{\rm el-ph1}}$ changes in time, see Fig.~\ref{effectiveT}(b).
For $g_3=0.05$, the reversal of relevant temperatures does not happen, see Figs.~\ref{effectiveT}(c) and (d).
Thus, the existence of the reversal behavior consistently explains the backward energy flow discussed in Fig.~\ref{energyflowanalysis}.

Now, we discuss the behavior of the effective temperature in detail. 
Firstly, reflecting that the effective temperature of $f^G$, i.e.,  $T^{G}$, is well defined, it well matches that of $\Pi_{\rm el-ph1}$, i.e., $T^{\Pi_{\rm el-ph1}}$. 
Note that we have $T^{G}=T^{\Pi_{\rm el-ph1}}$ in the MTM, which assumes the quasiparticle and qusiequilibrium approximations, see Appendix~\ref{app:GFtoBE}. 
Furthermore, the effective temperature of $\Sigma_{\rm el-ph1}$, i.e., $T^{\Sigma_{\rm el-ph1}}$, behaves in a way that is roughly located between $T^{G}$ and $T^{D_1}$.
Thus, the timing of the switch of magnitude relation between $T^{G}$ and $T^{\Sigma_{\rm el-ph1}}$  and that between $T^{D_1}$ and $T^{\Pi_{\rm el-ph1}}$ can be attributed to that of $T^{G}$ and $T^{D_1}$. 
These results suggests that the interpretation of the backward energy flow based on the MTM, where the direction of the energy flow is determined by the magnitude relation of effective temperatures, is reasonable in the weak electron-phonon coupling regime, although the phonon temperature is not fully well-defined, see Fig.~\ref{noneqdistribution} (see the additional discussion in Appendix~\ref{app:phononT_discussion}). 
The effective temperature $T^{\Sigma_{\rm el-ph2}}$ is smaller than $T^G$, so the electron energy always flows to phonon-2.
Similarly, $T^{\Pi_{\rm ph1-ph2}}$ which is equal to the equilibrium temperature $T_{\rm eq.}$ is smaller than $T^{D_1}$, hence the phonon-1 energy always flows to the phonon-2.

\subsection{Energy flows in intermediate electron-phonon coupling regime}\label{results:comparison}
\begin{figure}
\centering
\includegraphics[scale=0.45]{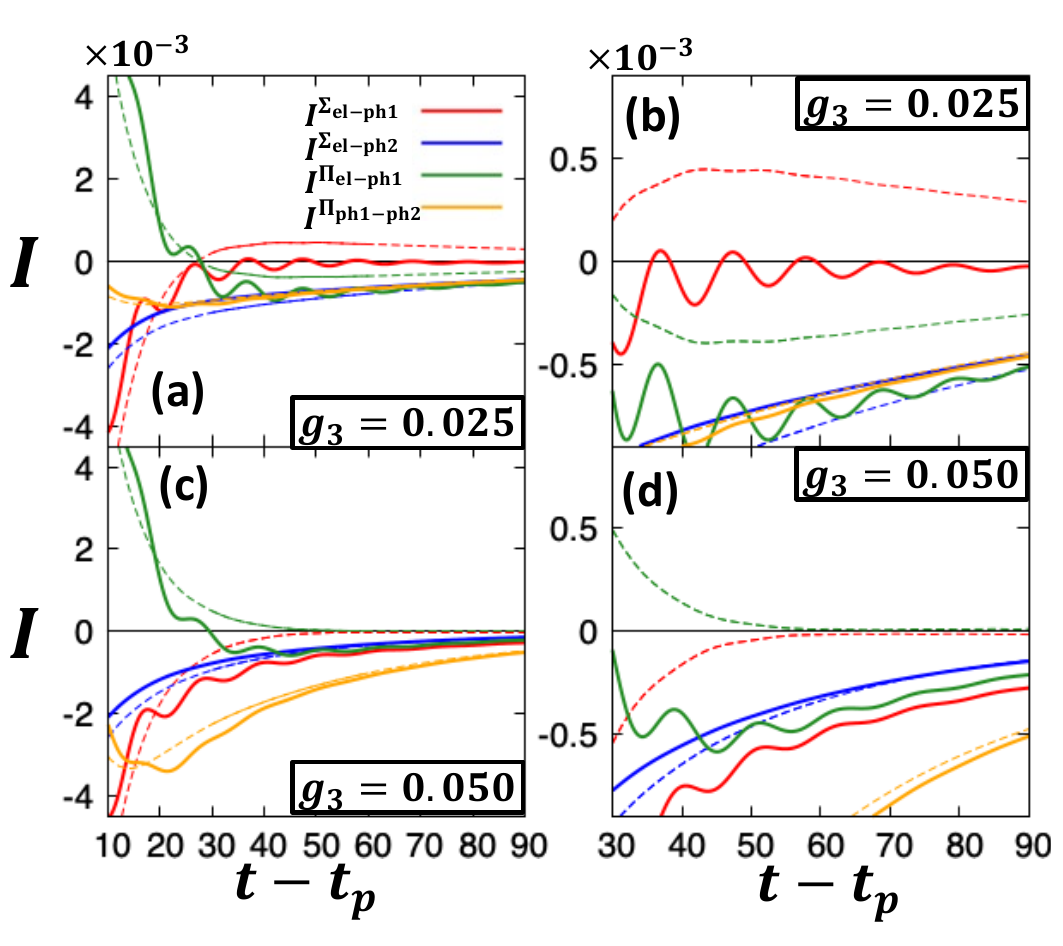}
\caption{
(Color online) 
(a) Energy flows for $g_1=0.35$, $g_2=0.2$, and $g_3=0.025$. (b) is enlarged view of (a).
(c) Energy flows for $g_1=0.35$, $g_2=0.2$, and $g_3=0.05$. (d) is enlarged view of (c).
We set $\omega_1=0.4$, and $\beta=20$.
For the bath parameters, we set $\omega_D=0.4$ and $\gamma=0.1$.
For the electric field, we set $E_{\rm el}=0.4$, $\Omega_{\rm el}=0.4$, and $\sigma=3$.
Solid lines represent $I^\Sigma$ and $I^\Pi$ expressed as Eqs.~(\ref{eq:Iph1el})-(\ref{eq:Iph2ph1}), and dashed lines represent the approximated values of those expressed as Eqs.~(\ref{eq:Ielapprox}) and (\ref{eq:Iph1approx}).
}
\label{energyflow2}
\end{figure}
\begin{figure*}
\centering
\includegraphics[scale=0.5]{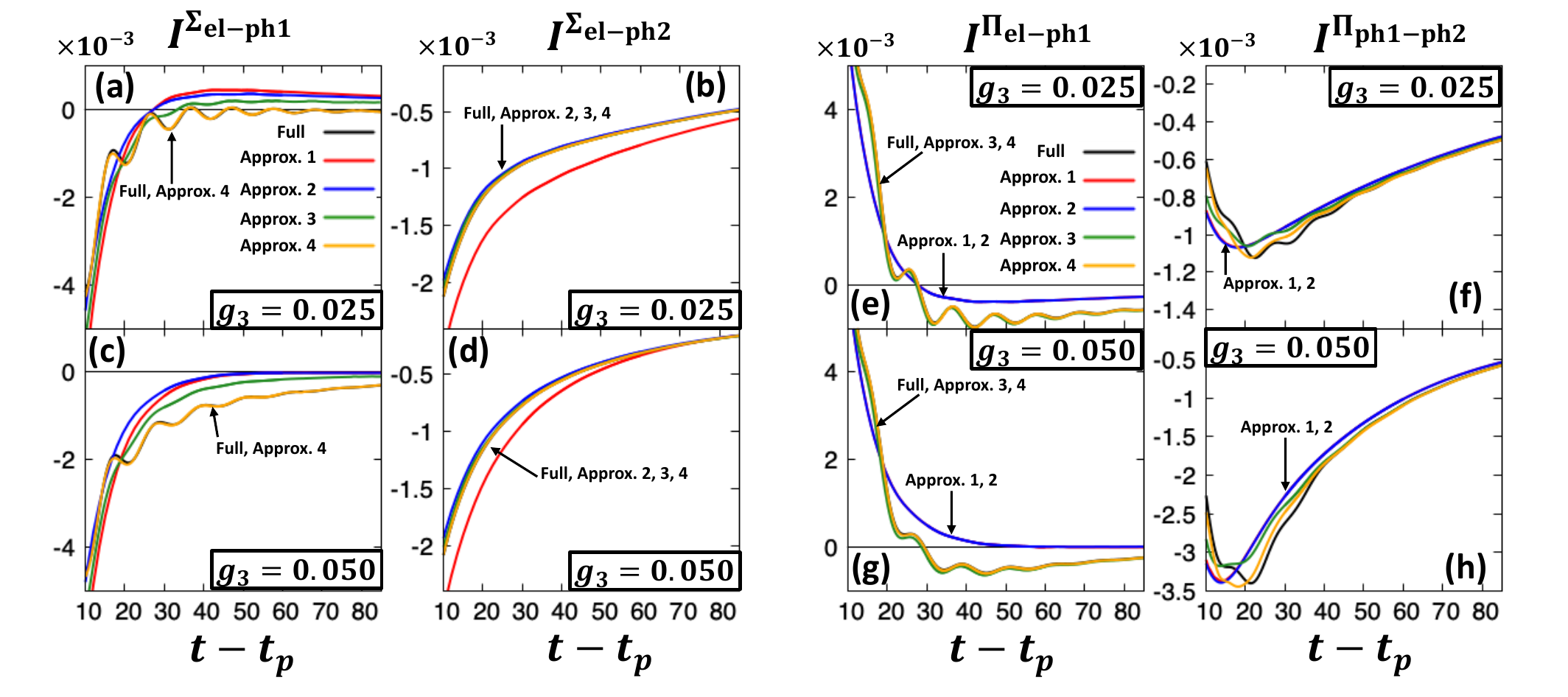}
\caption{
(Color online) 
Comparison of the full energy flows and the approximated energy flows listed in tables~\ref{table:electron_approximatedflow} and \ref{table:phonon_approximatedflow} for $g_1=0.35$ and $g_2=0.2$.
We set $\omega_1=0.4$, and $\beta=20$.
For the bath parameters, we set $\omega_D=0.4$ and $\gamma=0.1$.
For the electric field, we set $E_{\rm el}=0.4$, $\Omega_{\rm el}=0.4$, and $\sigma=3$.
Comparison of the full and approximated $I^{{\Sigma}_{\rm el-ph1}}$ for (a) $g_3=0.025$ and (c) $g_3=0.05$.
Comparison of the full and approximated $I^{{\Sigma}_{\rm el-ph2}}$ for (b) $g_3=0.025$ and (d) $g_3=0.05$.
Comparison of the full and approximated $I^{{\Pi}_{\rm el-ph1}}$ for (e) $g_3=0.025$ and (g) $g_3=0.05$.
Comparison of the full and approximated $I^{{\Pi}_{\rm ph1-ph2}}$ for (f) $g_3=0.025$ and (h) $g_3=0.05$.
}
\label{energyflow_compare}
\end{figure*}
\begin{table*}
\begin{tabular}{|c||c| }
\hline
Energy flow & $I^\Sigma$ \\
\hline 
\hline
Full & 
$-4{\rm Re}\left\{ \left[ \Sigma \ast G \ast \Delta \right]^< (t,t) \right\}$ 
\\ \hline
Approx. 1 &
Eq.~\eqref{eq:Ielapprox}
\\ \hline
Approx. 2 &
Eq.~\eqref{eq:Ielapproxnoquasi}
\\ \hline
Approx. 3 &
$-4\int_{-\infty}^{\infty}\frac{d\omega}{2\pi}\left\{{\rm Im}\left[\Delta\ast G\right]^R(\omega,t){\rm Im}\Sigma^<(\omega,t)-{\rm Im}\left[\Delta\ast G\right]^<(\omega,t){\rm Im}\Sigma^R(\omega,t)\right\}$
\\ \hline
Approx. 4 &
\begin{tabular}{c}
Eq.~\eqref{eq:Ielapprox_hyb}
\end{tabular} 
\\ \hline
\end{tabular}
\caption{Full and approximated electron energy flows}
\label{table:electron_approximatedflow}
\end{table*}

\begin{table*}
\begin{tabular}{|c||c| }
\hline 
Energy flow & $I^\Pi$ \\
\hline \hline
Full & 
${\rm Im}\left\{ \frac{\partial}{\partial t} \left[D_1\ast \Pi \right]^<(t,t')|_{t'=t}\right\}$
\\ \hline
Approx. 1 &
Eq.~\eqref{eq:Iph1approx}
\\ \hline
Approx. 2 &
$\frac{\omega_1}{2\pi}\int_{-\infty}^{\infty}\left\{\left({\rm Re}D_{d_1}^R(\omega,t)\right)\left({\rm Im}\Pi^<(\omega,t)\right)-\left({\rm Im}\Pi^R(\omega,t)\right)\left({\rm Re}D_{d_1}^<(\omega,t)\right) \right\}d\omega$
\\ \hline
Approx. 3 &
$\frac{\omega_1}{2\pi}\int_{-\infty}^{\infty}\left\{
\left({\rm Re}D_{d_1}^R(\omega,t)\right)\left({\rm Im}\Pi^<(\omega,t)\right)-\left({\rm Im}\Pi^R(\omega,t)\right)\left({\rm Re}D_{d_1}^<(\omega,t)\right) +\left({\rm Re}\Pi^R(\omega,t)\right)\left({\rm Im}D_{d_1}^<(\omega,t)\right)\right\}d\omega$
\\ \hline
Approx. 4 &
Eq.~\eqref{eq:Iph1_Dd1}
\\ \hline
\end{tabular}
\caption{Full and approximated phonon-1 energy flows}
\label{table:phonon_approximatedflow}
\end{table*}

Now, we study the energy flows in the stronger coupling regime and discuss how reasonable the approximated energy-flow equations ~(\ref{eq:Ielapprox}) and (\ref{eq:Iph1approx}) are.
Remember that these approximated equations are based on the gradient approximation and the quasiparticle approximations, both of which can break in the strong electron-phonon coupling regime.
As we see below, indeed, the equations~\eqref{eq:Ielapprox} and \eqref{eq:Iph1approx} fail to describe even the qualitative behavior of the energy flow.
By comparing the approximated energy flow equations of different levels, we discuss which approximations fail.
In this subsection, we fix $g_1=0.35$ and $g_2=0.2$, and change the phonon-phonon coupling $g_3$.
In these parameters, the coupling strength between electrons and two phonon modes is $\lambda\approx0.41 \ (\lambda_1\approx 0.358,\ \lambda_2\approx 0.048)$ for $g_3=0.025$ and 
$\lambda\approx0.44 \ (\lambda_1\approx 0.392,\ \lambda_2\approx 0.048)$ for $g_3=0.05$.
In these intermediate coupling regime, the Migdal approximation remains reasonable.

First, we compare the full energy flows (Eqs.~\eqref{eq:Iph1el}-\eqref{eq:Iph2ph1}) and their approximations with Eqs.~(\ref{eq:Ielapprox}) and (\ref{eq:Iph1approx}) in Figs.~\ref{energyflow2} (a)-(d).
Unlike in  the case for $g_1=g_2=0.2$ (Sec.~\ref{results:backflow}), the behavior of the full energy flow (solid line) and that of the approximated energy flow (dashed line) are qualitatively different.
In particular, the difference in  the behavior of $I^{\Sigma_{\rm el-ph1}}$ (red line) and $I^{\Pi_{\rm el-ph1}}$ (green line) is prominent.
For example, for $g_3=0.025$, the approximated $I^{\Sigma_{\rm el-ph1}}$ (red dashed line) clearly shows the sign change(the backward energy flow), while the full energy flow $I^{\Sigma_{\rm el-ph1}}$ (red solid line) does not show the sign change but converges to zero, see Fig.~\ref{energyflow2}(b).
For $g_3=0.05$, the approximated $I^{\Sigma_{\rm el-ph1}}$ (red dashed line) and $I^{\Pi_{\rm el-ph1}}$ (green dashed line) do not show the sign change, but the sign of the full energy flow $I^{\Pi_{\rm el-ph1}}$ (green solid line) changes from positive to negative, see Fig.~\ref{energyflow2}(d).
These results demonstrate that, if the electron-phonon coupling is strong, we cannot explain the origin of energy flows (in particular, their direction) from the difference between the nonequilibrium distribution functions of the Green's functions and the self-energies.

In order to identify the origin of the difference between the full and approximated energy flows, 
we compare the approximated energy flows in different levels.
For the electron energy flow $I^\Sigma$, in order to derive Eq.~\eqref{eq:Ielapprox} from Eq.~\eqref{eq:Ielapprox_Gk}, we use two approximations, the gradient approximation and the quasiparticle approximation.
To check the validity of these approximations, we introduce the approximated energy flows from Eq.~\eqref{eq:Ielapprox_Gk} with less approximations.
First, we introduce the approximated $I^\Sigma$ without the quasiparticle approximation (see Eq.~\eqref{eq:fGapprox}), but we approximate the electron spectrum with the Cauthy-Lorentz function
as 
\begin{align}
-\frac{1}{\pi}{\rm Im}G_k^R(\omega,t)\approx-\frac{1}{\pi}
\frac{{\rm Im}\Sigma_{\rm el}^R(\omega,t)}{\left(\omega-\epsilon_k-{\rm Re}\Sigma_{\rm el}^R(\omega,t)\right)^2+\left({\rm Im}\Sigma_{\rm el}^R(\omega,t)\right)^2}. \label{eq:G_approx}
\end{align}
This approximation corresponds to the approximated Dyson equation $G_k(\omega,t) \approx G_k^0(\omega,t)+G_k^0(\omega,t)\Sigma_{\rm el}(\omega,t) G_k(\omega,t)$. Thus, we use the gradient approximation partially.
This expression partially takes into account the renormalization of the electron spectrum due to the self-energy.
For simplicity, we approximate the nonequilibrium distribution function of $G_k$ by $f^G$.
Using these approximations, we can approximate $I^{\Sigma}$ as 
\begin{widetext}
\begin{align}
I^{\Sigma} &\approx
-8\int_{-\infty}^{\infty}d\epsilon \ \epsilon \rho(\epsilon)
\int_{-\infty}^{\infty}\frac{d\omega}{2\pi} 
\left\{
\frac{{\rm Im}\Sigma_{\rm el}^R(\omega,t)}{\left(\omega-\epsilon-{\rm Re}\Sigma_{\rm el}^R(\omega,t)\right)^2+\left({\rm Im}\Sigma_{\rm el}^R(\omega,t)\right)^2}
\right\}
{\rm Im}\Sigma^R(\omega,t)\left\{f^G(\omega,t)-f^\Sigma(\omega,t)\right\}. 
\label{eq:Ielapproxnoquasi}
\end{align}
\end{widetext}

Next, we introduce the expressions of energy flows making use of the DMFT.
In the DMFT, we can express $\frac{1}{N}\sum_k \epsilon_k G_k = [\Delta*G]$ using the hybridization function $\Delta$, see Appendix~\ref{app:derivation_of_Ielapprox_hyb} for details.
This allows us to rewrite Eq.~(\ref{eq:Ielapprox_Gk}) with $\Delta$ and without  the Green's function $G_k$ for each label $k$.
Then, the expression of $I^{\Sigma}$ is expressed as 
\begin{widetext}
\begin{align}
I^{\Sigma}\approx 
&-4\int_{-\infty}^{\infty}\frac{d\omega}{2\pi}
\left\{
{\rm Im}\left[\Delta\ast G\right]^R(\omega,t){\rm Im}\Sigma^<(\omega,t)-{\rm Im}\left[\Delta\ast G\right]^<(\omega,t){\rm Im}\Sigma^R(\omega,t)
\right\} \nonumber \\
&+
2\int_{-\infty}^{\infty}\frac{d\omega}{2\pi}
\left[
\left\{{\rm Re}\Sigma^R(\omega,t), {\rm Im}\left[\Delta\ast G\right]^<(\omega,t)\right\}_p
 -\left\{{\rm Re}\left[\Delta\ast G\right]^R(\omega,t), {\rm Im}\Sigma^<(\omega,t)\right\}_p
 \right]. \label{eq:Ielapprox_hyb}
\end{align}
\end{widetext}
Note that this expression is essentially the same as Eq.~\eqref{eq:Ielapprox_Gk} and free from the gradient approximation and the quasiparticle approximation.
 The gradient approximation corresponds to neglecting the second line of this expression (the Poisson bracket terms).
 
We summarize the expressions of the energy flow $I^{\Sigma}$ in table~\ref{table:electron_approximatedflow}.
Approx.~1 is the expression derived both with the gradient and quasiparticle approximations.
In Approx.~2 and Approx.~3, we use the gradient approximation, but do not ignore the renormalization of the spectrum of electrons due to the self-energy.
In Approx.~2, we approximate the spectrum of electrons as Eq.~\eqref{eq:G_approx} and $f^{G_k}$ as $f^G$.
Approx.~2 can be regarded as an extension of Approx.~1.
On the other hand, in Approx.~3, we do not use the above approximations used in Approx.~2.
In Approx.~4, we do not use both the gradient and quasiparticle approximations, although it already uses the gradient expansion.

For the phonon energy flow $I^\Pi$, in order to derive Eq.~\eqref{eq:Iph1approx} from Eq.~\eqref{eq:Iph1_Dd1}, we use the gradient approximation.
To clarify the validity of these approximations, we compare the results of different expressions of the energy flow $I^{\Pi}$, which are summarized in table~\ref{table:phonon_approximatedflow}.
Both Approx.2 and Approx. 3 use the gradient approximation to ignore the second line of Eq.~\eqref{eq:Iph1_Dd1}.
In Approx. 2, we further use Eq.~\eqref{eq:Dd1approx2}, where only the real prat of $D_{d_1}^<$ is considered.
We also note that Eq.~\eqref{eq:Iph1_Dd1} (Approx. 4) is based on the gradient expansion.

In Fig.~\ref{energyflow_compare}, we show the full and approximated energy flows listed up in tables~\ref{table:electron_approximatedflow} and \ref{table:phonon_approximatedflow}.
As for $I^{\Sigma_{\rm el-ph1}}$, there is no large difference between Approx.~1 (red line) and Approx.~2 (blue line), 
and these two approximated energy flows do not reproduce the behavior of the full energy flow (black line) both for $g_3=0.025$ and $g_3=0.05$, see Figs.~\ref{energyflow_compare}(a) and (c).
Furthermore, the energy flow from Approx.~3 slightly approaches the full energy flow compared to Approx.~1 and 2, but it still far from the full one.
In particular, Approx.~3 still shows a sign change, but the full energy flow converges to zero for $g_3=0.025$.
These comparisons suggest that the difference between the full energy flow and that approximated by Eq.~\eqref{eq:Ielapprox} is not solely attributed to the quasiparticle approximation.
On the other hand, Approx.~4 reproduce the full energy flow quantitatively.
This indicates that the gradient approximation is too rough in the current coupling regime, but the lowest-order correction in the gradient expansion is enough to reproduce the full flow.
As for $I^{\Sigma_{\rm el-ph2}}$, Approx.~2 already reproduces the full flow quantitatively well, see Figs.~\ref{energyflow_compare}(b) and (d).

As for the energy flow relevant to phonons, 
both of $I^{\Pi_{\rm el-ph1}}$ and $I^{\Pi_{\rm ph1-ph2}}$ evaluated from Approx.~1 (red line) and Approx.~2 (blue line) match well quantitatively for $g_3=0.025$ and $g_3=0.05$.
This means that approximations: ${\rm Re}D_{d_1}^R(\omega,t)\approx \omega {\rm Im}D_1^R(\omega,t)/\omega_1$,
${\rm Re}D_{d_1}^<(\omega,t)\approx \omega {\rm Im}D_1^<(\omega,t)/\omega_1$ are valid.
However, especially in $I^{\Pi_{\rm el-ph1}}$, these two approximated energy flows do not reproduce the full energy flow (black line).
Approx.~3 (green line) reproduces the full energy flow well, see Figs.~\ref{energyflow_compare}(e) and (g). 
This indicates that ${\rm Im}D_{d_1}^<$ plays an essential role, while the role of the gradient approximation is less important.

The above detailed analysis of the energy flows indicates that in the stronger electron-phonon coupling regime, the energy flows simply approximated with the difference of the nonequilibrium distributions are not applicable to discuss the qualitative feature of the energy flows. 
Still, we showed that the lowest order correction in the gradient expansion is enough to capture the correct behavior.

\section{SUMMARY AND OUTLOOK}\label{sec:summary}
In this work, we studied the energy flows in the relaxation process
in an extended Holstein model where electrons are coupled to two phonon modes.
Using two types of the Migdal approximation combined with the DMFT, we treated one phonon mode as the hot-phonon and the other mode as the heat-bath phonon.
Evaluating the energy flow between subsystems, we confirmed that the backward energy flow occurs in the relaxation process.
Focusing on this backward energy flow, we compared the results from several approximated expressions of the energy flows to reveal the validity of various approximations.
The simplest and intuitive expressions are those based on the gradient and quasiparticle approximations Eqs.~\eqref{eq:Ielapprox} and \eqref{eq:Iph1approx}, where the energy flows are directly associated with the difference of the nonequilibrium distributions of 
a relevant Green's function and self-energy. 
With a few more approximations, these expressions lead to the BE or the MTM, which were used to provide intuitive explanations of the backward energy flow.
Namely, the MTM tells that the direction of the energy flow is determined by the magnitude relation between the electron and phonon effective temperatures, and the backward energy flow originates from the reversal of this relation.

In our study, by comparing the full energy flow and the approximated energy flow, we found the simple expressions associated with the nonequilibrium distributions well explain the full energy flow including the backward energy flow in the weak electron-phonon coupling regime.
Thus, the backward energy flow is attributed to the reversal of the magnitude relation of relevant distributions.
Furthermore, we evaluated the effective temperatures from the distribution functions and confirmed that the picture provided by the MTM works reasonably well.
On the other hand, in the stronger electron-phonon coupling regime, the simplest expressions fail to reproduce the qualitative behavior of the energy flow,
which make the interpretation of the energy flow based on  the nonequilibrium distribution function  difficult.
By comparing the results with different levels of approximations, we discussed the importance of the gradient approximation and/or the quasiparticle approximation.
In particular, we found that the leading order correction of the gradient expansion is enough to reproduce the behavior of both energy flows relevant for electrons and phonons.

In this study, we simplified the setup, for example, by treating cold-phonons as heat-bath phonons, ignoring the phonon energy dispersion (the momentum-dependent frequency), and treating the phonon-phonon coupling as the two-phonon process.
However, for real materials, it is important to consider acoustic phonons with the energy dispersion~\cite{PhysRevLett.127.036402} and three-phonon process due to anharmonic phonon-phonon couplings~\cite{PhysRevB.96.174439,PhysRevB.97.054310,PhysRevB.102.184307}. 
By incorporating these setups, we can obtain a deeper understanding of the relaxation dynamics involving multiple electron-phonon couplings.
Another interesting direction is to study the relaxation dynamics under the direct photo-excitation of phonon modes.
A recent experimental study~\cite{PhysRevLett.126.077401} has demonstrated the suppression of electron-relaxation under the direct excitation of optical phonon modes by the terahertz light in a semiconductor, and it is expected to enable the control of electron dynamics via the phonon-excitation.
We expect to be able to address these issues by extending the model used here to the two-band system.
These are now under consideration.

\begin{acknowledgements}
Numerical calculations were performed with NESSi~\cite{SCHULER2020107484}.
This work is supported by Grant-in-Aid for Scientific Research from JSPS, KAKENHI Grant Nos. JP23KJ0883 (K. I.), JP22K03525, JP21H01025, JP19H05821 (A. K.), JP20K14412, JP21H05017 (Y. M.), 
JST, the establishment of university fellowships towards the creation of science technology innovation Grant No. JPMJFS2112 (K. I.), and JST CREST Grant No. JPMJCR1901 (Y. M.).
\end{acknowledgements}

\appendix
\section{Interaction energies and their flow equations} \label{app:interaction}
The electron-phonon-1 , electron-phonon-2 , and phonon-phonon interaction energies are respectively expressed as
\begin{align}
E_{\rm el-ph1}(t)=-2i\left[\Sigma_{\rm el-ph1} \ast G \right]^<(t,t), \\
E_{\rm el-ph2}(t)=-2i\left[\Sigma_{\rm el-ph2} \ast G \right]^<(t,t), \\
E_{\rm ph1-ph2}(t)=i\left[\Pi_{\rm ph1-ph2} \ast D_1\right]^<(t,t).
\end{align}
The flow equations of these energies are expressed as
\begin{widetext}
\begin{align}
\frac{d E_{\rm el-ph1}(t)}{dt}
&=-I^{\Pi_{\rm el-ph1}}-I^{\Sigma_{\rm el-ph1}}-4{\rm Re}\left\{\left[\Sigma_{\rm el} \ast G \ast \Sigma_{\rm el-ph1}\right]^<(t,t)\right\}, \\
\frac{d E_{\rm el-ph2}(t)}{dt}
&=-{\rm Im}\left\{\left. \frac{\partial}{\partial t}\left[\left\{\Sigma_p D_{2,p}^0\right\}\ast \left\{\left(\frac{g_2}{g_1}\right)^2\Pi_{\rm el-ph1}\right\} \right]^<(t,t')\right|_{t'=t}\right\}-I^{\Sigma_{\rm el-ph2}}-4{\rm Re}\left\{\left[\Sigma_{\rm el} \ast G \ast \Sigma_{\rm el-ph2}\right]^<(t,t)\right\}, \\
\frac{d E_{\rm ph1-ph2}(t)}{dt}
&=-{\rm Im}\left\{\left. \frac{\partial}{\partial t}\left[\Pi_{\rm ph1-ph2}\ast D_1\right]^<(t,t') \right|_{t'=t}\right\}-I^{\Pi_{\rm ph1-ph2}}.
\end{align}
\end{widetext}

In the normal Holstein model ($g_2=g_3=0$), $\Sigma_{\rm el}=\Sigma_{\rm el-ph1}$ and $-4{\rm Re}\left\{\left[\Sigma_{\rm el-ph1} \ast G \ast \Sigma_{\rm el-ph1}\right]^<(t,t)\right\}$=0.
Therefore, the magnitude difference of $I^{\Sigma_{\rm el-ph1}}$ and $I^{\Pi_{\rm el-ph1}}$ causes the time-evolution of $E_{\rm el-ph1}$ (this statement is equivalent to the energy conservation).
In our model,  $\frac{d E_{\rm el-ph1}(t)}{dt}$ includes the additional term, $-4{\rm Re}\left\{\left[\Sigma_{\rm el-ph2} \ast G \ast \Sigma_{\rm el-ph1}\right]^<(t,t)\right\}$, but this term cancels out $-4{\rm Re}\left\{\left[\Sigma_{\rm el-ph1} \ast G \ast \Sigma_{\rm el-ph2}\right]^<(t,t)\right\}$ included in $\frac{d E_{\rm el-ph2}(t)}{dt}$.

In this study, we treat the phonon-2 as the heat-bath (we do not consider the time-evolution of the phonon-2 energy), so there are no terms that cancel the first terms of above $\frac{d E_{\rm el-ph2}(t)}{dt}$ and $\frac{d E_{\rm ph1-ph2}(t)}{dt}$, and the total energy is not conserved.

\section{Relation with Boltzmann equation and Temperature model} \label{app:GFtoBE}
In this section, referring Ref.~\cite{FREERICKS2021147104}, we explain how the approximated energy flow equations~\eqref{eq:Ielapprox} and \eqref{eq:Iph1approx} are related to the BE and the MTM.
For simplicity, in this section, we consider the normal Holstein model ($g_2=g_3=0$) and derive the kinetic equations of energies in the BE and the 2TM using Eqs.~\eqref{eq:Ielapprox} and \eqref{eq:Iph1approx}.
Other derivations of the BE are discussed in detail, e.g., in Kamenev's textbook~\cite{kamenev_2023} and in previous studies~\cite{PhysRevB.98.134312,PhysRevB.104.085108}.

The BE is obtained by combining Eqs.~\eqref{eq:Ielapprox}, \eqref{eq:Iph1approx} and the self-energies approximated with the quasiparticle approximation; $-{\rm Im}G_k^R(\omega,t)/\pi \to \delta(\omega-\epsilon_k)$ and $-{\rm Im}D_1^R(\omega,t)/\pi \to \delta(\omega-\omega_1)-\delta(\omega+\omega_1)$. 
With the approximation, the imaginary parts of lesser and retarded components of the electron self-energy are written as
\begin{widetext}
\begin{align}
{\rm Im}\Sigma_{\rm el-ph1}^<(\omega,t)
&\approx 2\pi g^2
\left\{
\rho(\omega-\omega_1)f^G(\omega-\omega_1,t)f^{D_1}(\omega_1,t)+\rho(\omega+\omega_1)f^G(\omega+\omega_1,t)\left(1+f^{D_1}(\omega_1,t)\right)
\right\}, \\
{\rm Im}\Sigma_{\rm el-ph1}^R(\omega,t)
&\approx-\pi g^2
\left[
\rho(\omega-\omega_1)\left\{f^G(\omega-\omega_1,t)f^{D_1}(\omega_1,t)+\left(1-f^G(\omega-\omega_1,t)\right)\left(1+f^{D_1}(\omega_1,t)\right)\right\}
\right. \nonumber \\
&\left.
+\rho(\omega+\omega_1)\left\{f^G(\omega+\omega_1,t)\left(1+f^{D_1}(\omega_1,t)\right)+\left(1-f^G(\omega+\omega_1,t)\right)f^{D_1}(\omega_1,t)\right\}
\right].
\end{align}
\end{widetext}
Here, we use Eq.~\eqref{eq:fGapprox}.
The retarded component involves all scattering processes associated with a phonon absorption and a  phonon emission.
On the other hand, the lesser component includes only processes where the electron energy becomes $\omega$ by absorbing or emitting a phonon.
Therefore, the self-energy distribution $f^{\Sigma_{\rm el-ph1}}=-{\rm Im}\Sigma_{\rm el-ph1}^</(2{\rm Im}\Sigma_{\rm el-ph1}^R)$ means the probability of the one-phonon absorption or emission process that makes the electron energy $\omega$ among all scattering processes.
Combining this expression of the self-energy and Eq.~\eqref{eq:Ielapprox},  we obtain
\begin{widetext}
\begin{align}
\frac{d E_{\rm kin}}{dt}&=
4\pi g^2 \int_{-\infty}^{\infty}d\omega \ \omega \rho(\omega)
\left[
\rho(\omega-\omega_1)
\left\{
\left(1-f^G(\omega,t)\right)f^G(\omega-\omega_1,t)f^{D_1}(\omega_1,t)-f^G(\omega,t)\left(1-f^G(\omega-\omega_1,t)\right)\left(1+f^{D_1}(\omega_1,t)\right)
\right\} \right. \nonumber \\
&\left.
+\rho(\omega+\omega_1)
\left\{
\left(1-f^G(\omega,t)\right)f^G(\omega+\omega_1,t)\left(1+f^{D_1}(\omega_1,t)\right)-f^G(\omega,t)\left(1-f^G(\omega+\omega_1,t)\right)f^{D_1}(\omega_1,t)
\right\}
\right].
\end{align}
\end{widetext}
This equals to the kinetic equation of the electron energy using the BE in the DMFT limit (without momentum conservation~\cite{PhysRevB.98.134312}).

Similarly, the imaginary parts of lesser and retarded components of the phonon-1 self-energy can be approximated as
\begin{widetext}
\begin{align}
{\rm Im}\Pi_{\rm el-ph1}^<(\omega,t)&\approx
-4\pi g^2 \int_{-\infty}^{\infty}\rho(\omega+\omega')f^G(\omega+\omega',t)\rho(\omega')\left(1-f^G(\omega',t)\right) d\omega', \\
{\rm Im}\Pi_{\rm el-ph1}^R(\omega,t)&\approx
-2\pi g^2 \int_{-\infty}^{\infty}\rho(\omega+\omega')\rho(\omega')
\left\{
f^G(\omega',t)\left(1-f^G(\omega+\omega',t)\right)-f^G(\omega+\omega',t)\left(1-f^G(\omega',t)\right)
\right\}d\omega'.
\end{align}
\end{widetext}
The retarded component involves the processes of absorbing and emitting energy $\omega$ throughout the coupling with electrons.
The lesser component only includes the process of emitting the energy $\omega$.
Therefore, the self-energy distribution $f^{\Pi_{\rm el-ph1}}={\rm Im}\Pi_{\rm el-ph1}^</(2{\rm Im}\Pi_{\rm el-ph1}^R)$ can be regarded as the probability of this energy-emission process among all scattering processes.
Using the expression of the self-energy and Eq.~\eqref{eq:Iph1approx}, we obtain
\begin{widetext}
\begin{align}
\frac{dE_{\rm ph1}}{dt}=
4\pi g^2 \omega_1\int_{-\infty}^{\infty}d\omega \; \rho(\omega)
\Bigl\{
f^G(\omega,t)\rho(\omega-\omega_1)\left(1-f^G(\omega-\omega_1,t)\right)\left(1+f^{D_1}(\omega_1,t)\right) \nonumber \\
-f^G(\omega,t)\rho(\omega+\omega_1)\left(1-f^G(\omega+\omega_1,t)\right)f^{D_1}(\omega_1,t)
\Bigl\}.
\end{align}
\end{widetext}
As in the case of the electron kinetic energy, this expression is identical to the kinetic equation of the phonon-1 energy in the BE.
Note that the above equations for $E_{\rm kin}$ and $E_{\rm ph1}$ satisfy the energy conservation, i.e., $\frac{dE_{\rm kin}}{dt}+\frac{dE_{\rm ph1}}{dt}=0$.

In order to derive the 2TM from the BE, we further apply the quasiequilibrium approximation.
Namely, we assume that the nonequilibrium distribution functions of electrons and phonons can be approximated with the Fermi-Dirac distribution or the Bose-Einstein distribution with time-dependent temperatures $T_{\rm el}(t)$ and $T_{\rm ph1}(t)$.
In this approximation, the nonequilibrium distribution function $f^{\Pi_{\rm el-ph1}}(\omega,t)$ matches $\frac{1}{\exp(\omega_1/T_{\rm el}(t))-1}$ and the kinetic equations of $E_{\rm kin}$ becomes
\begin{align}
\frac{dE_{\rm kin}}{dt}
&=-\frac{dE_{\rm ph1}}{dt} \nonumber \\
&=\frac{\omega_1}{\tau_{\rm ph1}(t)}\left\{\frac{1}{\exp(\omega_1/T_{\rm ph1}(t))-1}-\frac{1}{\exp(\omega_1/T_{\rm el}(t))-1}\right\} \nonumber \\
&\approx -\frac{C_{\rm ph1}(t)}{\tau_{\rm ph1}(t)}\left(T_{\rm el}(t)-T_{\rm ph1}(t)\right)\nonumber \\
&\equiv -G(t)\left(T_{\rm el}(t)-T_{\rm ph1}(t)\right).
\end{align}
Here, we use the Taylor expansion of the Bose-Einstein distribution.
We introduce $\frac{1}{2\tau_{\rm ph1}(t)}=-{\rm Im}\Pi^R(\omega_1,t)$, $C_{\rm ph1}(t)=\frac{\partial E_{\rm ph1}}{\partial T_{\rm ph1}}=\frac{\partial}{\partial T_{\rm ph1}}\frac{\omega_1}{\exp(\omega_1/T_{\rm ph1})-1}$, and $G(t)\equiv \frac{C_{\rm ph1}(t)}{\tau_{\rm ph1}(t)}$.
Using the specific heat of electrons $C_{\rm el}(t)=\frac{\partial E_{\rm el}}{\partial T_{\rm el}}$, we obtain the equation of the 2TM~\cite{PhysRevLett.59.1460,Mahan2000,PhysRevB.96.174439},
\begin{align}
C_{\rm el}(t)\frac{\partial T_{\rm el}(t)}{\partial t}=-C_{\rm ph1}(t)\frac{\partial T_{\rm ph1}(t)}{\partial t}
=-G(t)\left(T_{\rm el}(t)-T_{\rm ph1}(t)\right).
\end{align}

\section{Time-evolution of nonequilibrium spectral functions} \label{app:spectrum}
\begin{figure*}
\centering
\includegraphics[scale=1.2]{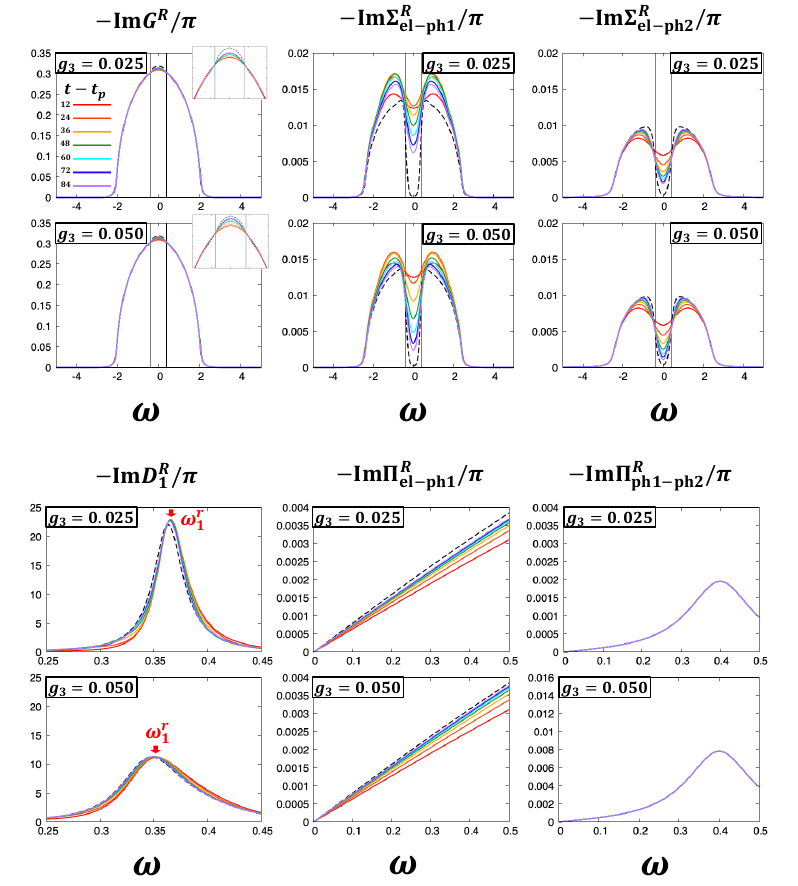}
\caption{
(Color online) Time-evolution of nonequilibrium spectral functions (Wigner transforms of the Green's functions and the self-energies) for $g_1=g_2=0.2$, $\omega_1=0.4$, and $\beta=20$.
For the bath parameters, we set $\omega_D=0.4$ and $\gamma=0.1$.
For the electric field, we set $E_{\rm el}=0.4$, $\Omega_{\rm el}=0.4$, and $\sigma=3$.
Small inset figures represent the enlarged view of the peak structures of $-{\rm Im}G^R/\pi$.
Black dashed lines show equilibrium values.
Black horizontal lines in panels for $-{\rm Im}G^R/\pi$, $-{\rm Im}\Sigma^R_{\rm el-ph1}/\pi$, and $-{\rm Im}\Sigma^R_{\rm el-ph2}/\pi$ represent the phonon window, i.e., $\omega \in [-\omega_1,\omega_1]$ ($\omega_1=\omega_D$).
}
\label{noneqspectrum}
\end{figure*}
In Fig.~\ref{noneqspectrum}, we show the time-evolution of the nonequilibrium spectral functions (the Wigner transforms of the Green's functions and the self-energies expressed as Eq.~\eqref{eq:Wigner}) for $g_1=g_2=0.2, \ g_3=0.025 \ {\rm and} \ 0.05$.
In equilibrium (black dashed line), spectrums of self-energies $\Sigma_{\rm el-ph1}$ and $\Sigma_{\rm el-ph2}$ have the relatively small value (valley structure) in the energy range $|\omega|<\omega_1=\omega_D=0.4$.
This means that the electron-like quasiparticle cannot decay by emitting the phonon in this energy range~\cite{Mahan2000,PhysRevB.91.045128,PhysRevB.98.245110}.
This characteristic structure of the imaginary part of the self-energy causes the peak structure of the real part of the self-energy around $\omega_1=\omega_D$ which is obtained using the Kramers-Kronig relation.
As a result, the peak structure of the electron spectrum ($-{\rm Im}G^R/\pi$) appears  in the energy range $|\omega|<\omega_1=\omega_D=0.4$ (see the insets of panels for $-{\rm Im}G^R/\pi$ in Fig.~\ref{noneqspectrum}).
If we set $\omega_D \neq \omega_1$, other characteristic structures of the electron spectrum appear around $\pm \omega_D$ as in Ref.~\cite{Covaci_2007}.
By the photo-excitation of electrons, the depth of valley structure in the self-energy spectrum becomes shallower and the peak structure of the electron spectrum is suppressed~\cite{PhysRevX.3.041033,PhysRevB.91.045128,PhysRevX.8.041009,PhysRevB.98.245110}.
This is consistent with that the electron temperature effectively increases (in equilibrium, if the temperature increases, similar phenomena can be observed).

The spectral function of phonon-1 ($-{\rm Im}D_1^R/\pi$) has the peak structure at the renormalized phonon-1 frequency $\omega_1^r$ different from the bare frequency $\omega_1=0.4$.
By the photo-excitation of electrons, the peak position slightly approaches to the bare frequency because the phonon-1 temperature effectively increases and the electron-phonon correlation is 
practically weaken~\cite{PhysRevB.91.045128,PhysRevB.98.245110}.
The consistent behavior can be seen in the imaginary part of the phonon-1 self-energy ($-{\rm Im}\Pi^R_{\rm el-ph1}/\pi$).
The photo-excitation causes this imaginary part to decrease from its equilibrium value.
This imaginary part of the self-energy is related to the width of the phonon-1 spectrum $-{\rm Im}D_1^R/\pi$ and this width decreases if the electron-phonon correlation is weaken.
In this work, the phonon-2 is treated as the heat-bath. Therefore, the self-energy ($-{\rm Im}\Pi^R_{\rm ph1-ph2}/\pi$) does not change from the equilibrium value.

\section{Oscillation of energy flow} \label{app:oscillation}
\begin{figure}
\centering
\includegraphics[scale=0.35]{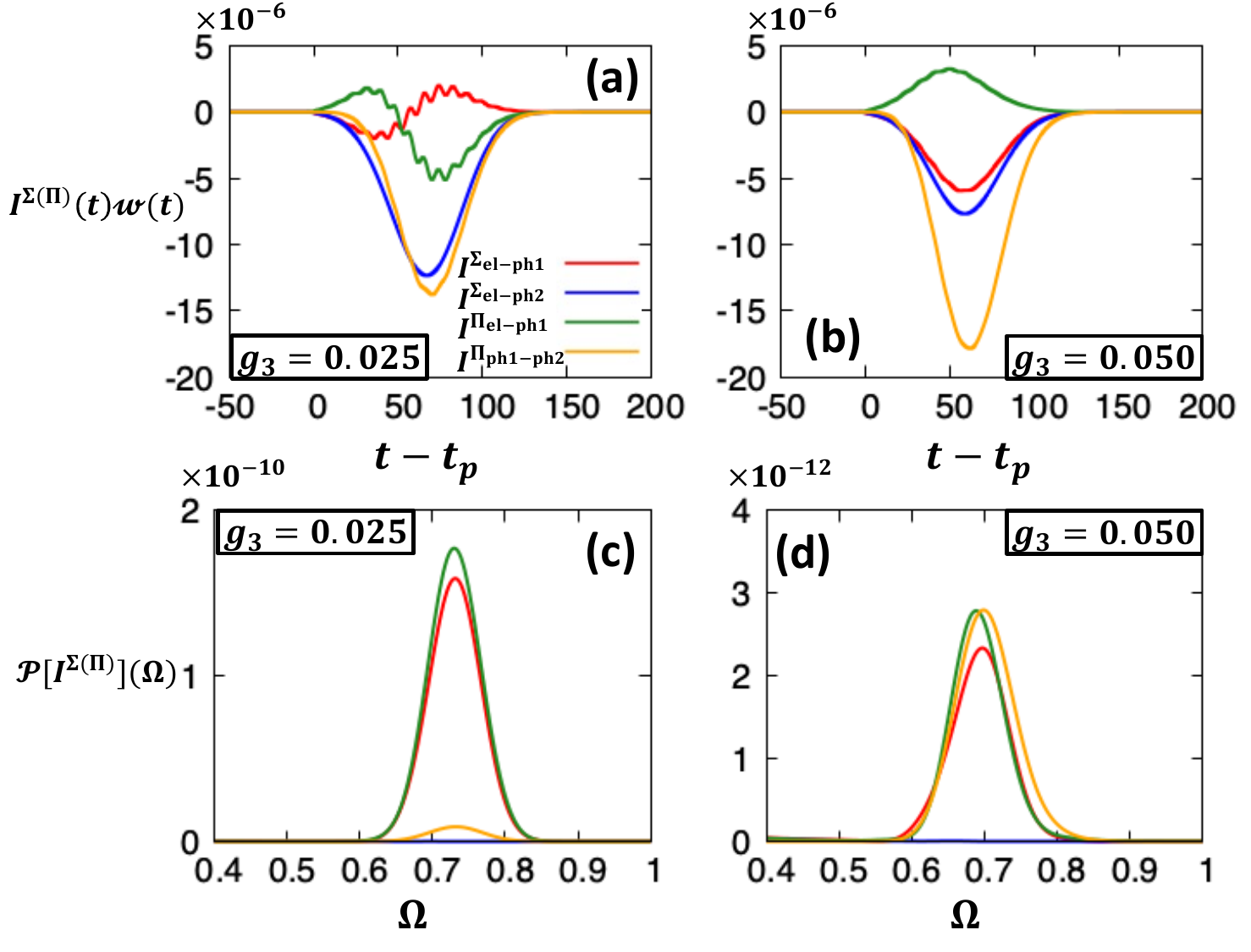}
\caption{
(Color online)
Product of the energy flow $I^{\Sigma(\Pi)}$ and the window function $\mathit{w}$ for (a) $g_1=g_2=0.2,\ g_3=0.025$ and (b) $g_1=g_2=0.2,\ g_3=0.05$.
Fourier power spectrum of the energy flow $I^{\Sigma(\Pi)}$ for (c) $g_1=g_2=0.2,\ g_3=0.025$ and (d) $g_1=g_2=0.2,\ g_3=0.05$.
We set $\omega_1=0.4$, and $\beta=20$.
For the bath parameters, we set $\omega_D=0.4$ and $\gamma=0.1$.
For the electric field, we set $E_{\rm el}=0.4$, $\Omega_{\rm el}=0.4$, and $\sigma=3$.
}
\label{power}
\end{figure}
In order to extract the oscillation frequency $\omega_{osc}$ of the energy flow shown in Fig.~\ref{energyflow},  we calculate the Fourier transform of the energy flow.
The Fourier transform of the energy flow $I$ and the power spectrum of it are defined as
\begin{align}
\mathcal{F}\left[I^{\Sigma(\Pi)}\right](\Omega)&=\int_{-\infty}^{\infty}I^{\Sigma(\Pi)}(t)\mathit{w}(t)e^{i\Omega t}dt, \\
\mathcal{P}\left[I^{\Sigma(\Pi)}\right](\Omega)&=\left| \mathcal{F}\left[I^{\Sigma(\Pi)}\right](\Omega) \right|^2.
\end{align}
Here, $\mathit{w}(t)$ is the window function and we adopt the following Gaussian function as $\mathit{w}(t)$ with two parameters $t_{\mu}$ and $t_\sigma$,
\begin{align}
\mathit{w}(t)=\frac{1}{\sqrt{2\pi}t_\sigma}{\rm exp}\left\{{-\frac{(t-t_{\mu})^2}{2t_\sigma^2}}\right\}.
\end{align}
We set $t_{\mu}-t_p=75$ and $t_\sigma=20$.
These parameters are selected to see a clear peak structure of $\mathcal{P}\left[I^{\Sigma(\Pi)}\right]$.
Figures~\ref{power} (a) and (b) show the product of $I^{\Sigma(\Pi)}$ and $\mathit{w}$ for $g_1=g_2=0.2,\ g_3=0.025 \ {\rm and} \ 0.05$.
We can see the fine oscillation on major curves.
Figures~\ref{power} (c) and (d) show the power spectrum $\mathcal{P}\left[I^{\Sigma(\Pi)}\right]$ for $g_1=g_2=0.2,\ g_3=0.025 \ {\rm and} \ 0.05$.
Power spectrums of three energy flow $I^{\Sigma_{\rm el-ph1}}$, $I^{\Pi_{\rm el-ph1}}$, and $I^{\Pi_{\rm ph1-ph2}}$ have the clear peak at $\omega\approx 0.73$ for $g_3=0.025$ and $\omega\approx 0.69\sim0.70$ for $g_3=0.05$ while we can hardly see the peak structure for $I^{\Sigma_{\rm el-ph2}}$.
These peak positions are about twice of the renormalized phonon-1 frequency $\omega_1^r$ (the peak position of $-{\rm Im}D^R_1/\pi$) (see Fig. \ref{noneqspectrum}).
Note that in the smaller $\omega$ region, the power spectrums have the main peaks (not shown).
However, these are the contribution from the major curves of $I^{\Sigma(\Pi)}(t)\mathit{w}(t)$, so differ from the oscillation related to the phonon-1 frequency.

\section{Discussion about effective temperatures} \label{app:phononT_discussion}
\begin{figure}
\centering
\includegraphics[scale=0.3]{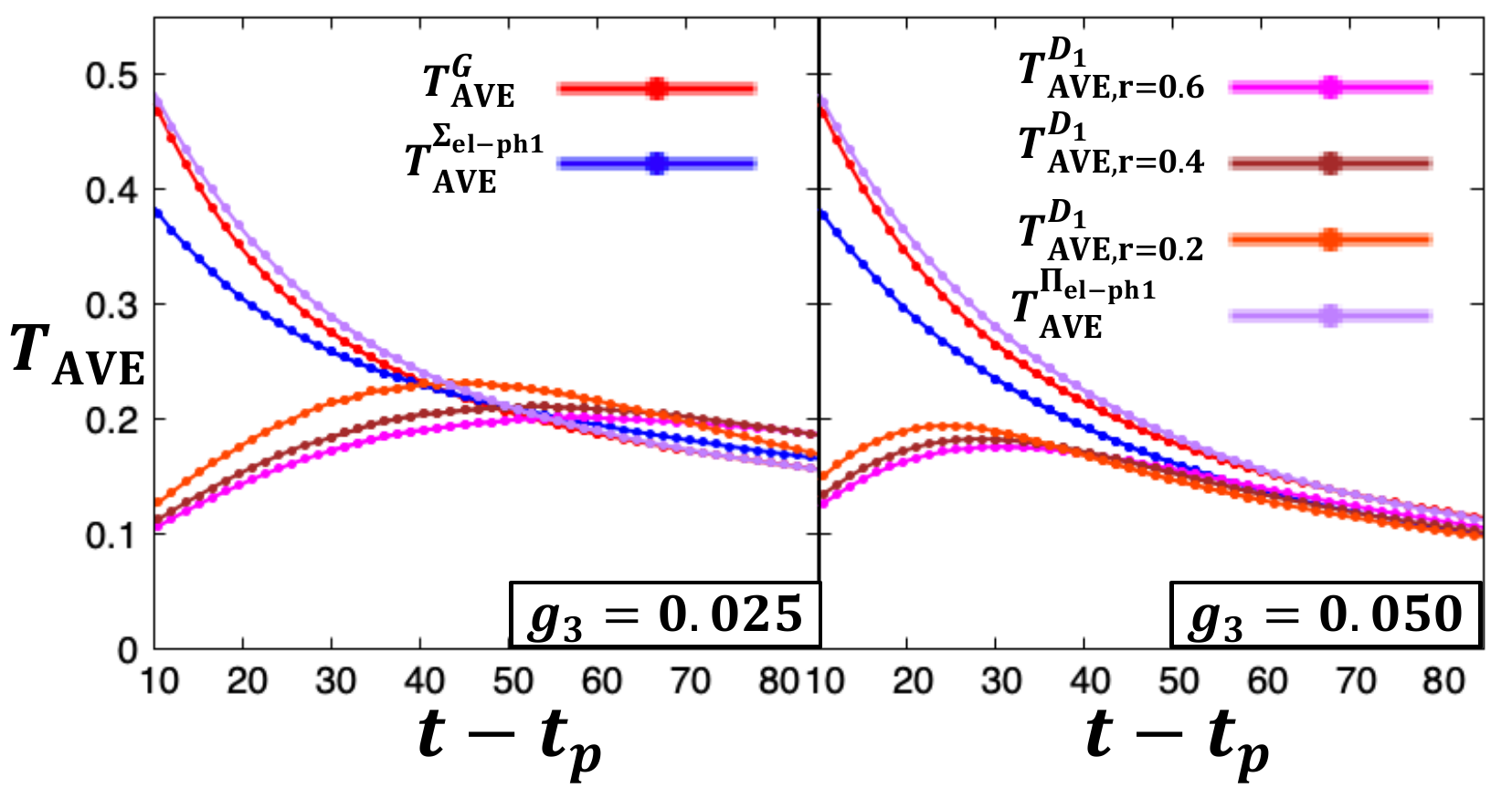}
\caption{
(Color online) Evolution of effective temperatures of the Green's functions and the self-energies expressed as Eq.~\eqref{eq:effectiveT}. 
We set $g_1=g_2=0.2$, $\omega_1=0.4$, and $\beta=20$.
For the bath parameters, we set $\omega_D=0.4$ and $\gamma=0.1$.
For the electric field, we set $E_{\rm el}=0.4$, $\Omega_{\rm el}=0.4$, and $\sigma=3$.
(a) Effective temperatures for $g_3=0.025$. (b) Effective temperatures for $g_3=0.05$.
$T^G_{\rm AVE}$, $T^{\Sigma_{\rm el-ph1}}_{\rm AVE}$, and $T^{\Pi_{\rm el-ph1}}_{\rm AVE}$ are the same as those plotted in Fig.~\ref{effectiveT}.
}
\label{phononTdiscussion}
\end{figure}
As we mentioned in Sec.~\ref{results:backflow}, the nonequilibrium distribution function $f^{D_1}$ deviates from the Bose-Einstein distribution.
Therefore, the average and standard deviation of effective temperature $T^{D_1}$  are affected by the fitting range $\mathbb{F}$.
In Fig.~\ref{phononTdiscussion}, we show $T^{D_1}_{\rm AVE,r}$ for several fitting ranges 
$\mathbb{F}_r=\left\{\omega \mid -{\rm Im}D_1^R(\omega,t)/\pi>r\times -{\rm Im}D_1^R(\omega_1^r(t),t)/\pi\right\}$ with $0<r<1$ in addition to $T^G_{\rm AVE}$, $T^{\Sigma_{\rm el-ph1}}_{\rm AVE}$, and $T^{\Pi_{\rm el-ph1}}_{\rm AVE}$.
The effective temperature $T^{D_1}_{\rm AVE,r=0.2}$ is the same as $T^{D_1}_{\rm AVE}$ plotted in Fig.~\ref{effectiveT}.
For $T^{\Pi_{\rm el-ph1}}_{\rm AVE}$, we plot the case for $r=0.2$ because $T^{\Pi_{\rm el-ph1}}_{\rm AVE}$ is mostly unaffected by the fitting range $\mathbb{F}_r$.
We can see that $T^{\Pi_{\rm el-ph1}}_{\rm AVE}$ well matches to $T^G_{\rm AVE}$ as described in Sec.~\ref{results:backflow}.

As one can see, the values of $T^{D_1}_{\rm AVE}$ depend on the fitting range $\mathbb{F}_r$ both for $g_3=0.025$ and $g_3=0.05$.
However, in any choice of $\mathbb{F}_r$, 
$T^{\Sigma_{\rm el-ph1}}_{\rm AVE}$ is roughly located between $T^{G}_{\rm AVE}$ and $T^{D_1}_{\rm AVE}$.
Thus, the choice of $\mathbb{F}_r$ hardly affects the discussion in terms of the effective temperatures in Sec.~\ref{results:backflow}.

\section{Derivation of Equation~\eqref{eq:Ielapprox_hyb}} \label{app:derivation_of_Ielapprox_hyb}
Within DMFT, the convolution of the hybridization function $\Delta$ and the lattice Green's function $G$ can be written as 
\begin{align}
\left[\Delta \ast G \right](t,t')=\frac{1}{N}\sum_{k(t)}\epsilon_{k(t)}(t)G_{k(t)}(t,t').
\end{align}
Here, $k(t)$ is the index of eigenstate for the free electron system.
In the Bethe lattice, when we consider the Peierls substitution, the index of eigenstate depends on time.
(In the periodic system where one can perform the Fourier transform, the index of eigenstate is the wave-number and this does not depend on time.)
Still, after the photo-excitation ($t \gg t_p$), the index of eigenstate does not depend on time, i.e., $\left\{k(t)\right\}=\left\{k\right\}$.

Now, we consider the Wigner transforms of the retarded and lesser components of $\left[\Delta \ast G \right]$ in $t_a \gg t_p$. They are expressed as 
\begin{align}
&\left[\Delta \ast G \right]^R(\omega,t_a)=\frac{1}{N}\sum_{k}\epsilon_{k}G_{k}^R(\omega,t_a), \\
&\left[\Delta \ast G \right]^<(\omega,t_a)\approx \frac{1}{N}\sum_{k}\epsilon_{k}G_{k}^<(\omega,t_a).
\end{align}
Note that the expression of the lesser component is the approximated one.
Remember that the Wigner transform includes the integration with respect to the relative time $t_r$, and for the lesser component the integration range may include the times during the pulse, where the indices $k$ and the energies $\epsilon_{k}$ of eigenstates depend on time. 
Still, since the lesser component of the Green's function converges to zero as $|t_r|$ increases, the contribution from the times under the pulse should be small for $t_a \gg t_p$.
On the other hand, the retarded part does not include the contribution from the times under the pulse for $t_a \gg t_p$.
Using above equations to Eq.~\eqref{eq:Ielapprox_Gk}, we can derive Eq.~\eqref{eq:Ielapprox_hyb}.

\bibliography{refs}

\end{document}